
\input harvmac
\def\sect#1\par{\par\ifdim\lastskip<\medskipamount
        \bigskip\medskip\goodbreak\else\nobreak\fi
        \noindent{\sectfont{#1}}\par\nobreak\medskip}
\def\frac#1#2{{\textstyle{#1\over#2}}}     
\def\half{{\textstyle{1\over{\raise.1ex\hbox{$\scriptstyle{2}$}}}}}
\def\on#1#2{{\buildrel{\mkern2.5mu#1\mkern-2.5mu}\over{#2}}}
\def\bop#1{\setbox0=\hbox{$#1M$}\mkern1.5mu
	\vbox{\hrule height0pt depth.04\ht0
	\hbox{\vrule width.04\ht0 height.9\ht0 \kern.9\ht0
	\vrule width.04\ht0}\hrule height.04\ht0}\mkern1.5mu}
\def\bo{{\mathpalette\bop{}}}                        
\def\dot#1{\on{\hbox{\bf .}}{#1}}                
\def\dzm{{\partial _-}}
\def\dzp{{\partial _+}}

\def\eps {{\epsilon}}
\def\a {{\alpha}}
\def\l {{\lambda}}
\def\P {{\Phi}}

\def\b {{\beta}}
\def\g {{\gamma}}
\def\d {{\delta}}
\def\s {{\sigma}}
\def\m {{\mu}}
\def\ah {{\alpha}}
\def\mh {{\mu}}
\def\bh {{\beta}}

\def\ad {{\dot\alpha}}
\def\md {{\dot\mu}}
\def\bd {{\dot\beta}}
\def\gd {{\dot\gamma}}

\def\adh {{{\dot\alpha}}}
\def\mdh {{{\dot\mu}}}
\def\bdh {{{\dot\beta}}}

\def\k {{\kappa}}

\def\t {{\theta}}
\def\tb {{\bar\theta}}
\def\ta {{\theta^\alpha}}
\def\tba {{\bar\theta^\ad}}
\def\th {{\hat\theta}}
\def\tbh {{\hat{\bar\theta}}}
\def\p {{\partial}}
\def\N {{\nabla}}
\def\Nb {{\bar\nabla}}
\def\Nh {{\hat{\nabla}}}
\def\Nbh {{\hat{\bar\nabla}}}

\def\L {{\Lambda}}
\def\Lb {{\bar\Lambda}}
\def\Lbh {{\hat{\bar \Lambda}}}
\def\Lh {{\hat{\Lambda}}}
\centerline{\titlefont Superspace Effective Actions for 4D
Compactifications}
\centerline{\titlefont of Heterotic and Type II Superstrings}
\bigskip\centerline{Nathan Berkovits}
\bigskip\centerline{Inst. de F\'{\i}sica, Univ. de S\~ao Paulo,
CP 66318, S\~ao Paulo, SP 05389-970, Brasil}
\centerline{and IMECC, Univ. de Campinas,
CP 1170, Campinas, SP 13100, Brasil}
\centerline{e-mail: nberkovi@snfma1.if.usp.br}
\bigskip\centerline{and}
\bigskip\centerline{Warren Siegel}
\bigskip\centerline{ITP, State Univ. of NY, Stony Brook, NY 11794-3840, USA}
\centerline{e-mail: siegel@insti.physics.sunysb.edu}
\vskip .2in
\centerline{\bf Abstract}

\narrower

Two-dimensional sigma models are defined for the new manifestly
spacetime supersymmetric description of four-dimensional compactified
superstrings.  The resulting target-superspace effective action is
constrained by the way the spacetime dilaton couples to the worldsheet
curvature:  For the heterotic superstring, the worldsheet curvature
couples to the real part of a chiral multiplet, and for Type II it couples to
the real part of the sum of a vector multiplet and a tensor hypermultiplet.

For the Type II superstring, this contradicts the standard folklore that
only a hypermultiplet counts string-loops, explains the peculiar dilaton
coupling of Ramond-Ramond fields, and allows the effective action to be
easily written in N=2 $4D$ superspace. It also implies that
vector multiplet interactions get no quantum corrections, while
hypermultiplet interactions can only get corrections if mirror
symmetry is non-perturbatively broken.

\Date{October 1995,  IFUSP-P-1180,  ITP-SB-95-41 }

\newsec{ Introduction}

There are two ways to construct low-energy effective actions in string theory.
One can define a two-dimensional sigma model for the string in a
curved background, use conformal invariance to determine the equations of
motion,
and look for an action which provides these equations of
motion\ref\cal{E. Fradkin and A. Tseytlin,
Phys. Lett. 158B (1985) 316\semi E. Fradkin and A. Tseytlin,
Nucl. Phys. B261 (1985) 1\semi
C. Callan, D. Friedan, E. Martinec, and M. Perry,
Nucl. Phys. B262 (1985) 593.}.
Alternatively, one can calculate on-shell S-matrix scattering amplitudes
and look for an action which yields these amplitudes\ref\bru
{J. Scherk, Nucl. Phys. B31
(1971) 222.}.

For the superstring
in the RNS formalism, both of these methods are made clumsy by the
complicated nature of the Ramond fields. For this reason, the part of the
effective action coming from the Ramond fields is much less understood
than the part coming from the Neveu-Schwarz sector.
In light of
recent conjectures relating non-perturbative states with
the Ramond-Ramond sector of the type II
superstring\ref\HT{C. Hull and P. Townsend, Nucl. Phys. B438 (1995) 109.},
 this lack of understanding is especially bothersome.
For example,
the $F^2$ term for the graviphoton field strength appears to be independent
of the dilaton $\varphi$, instead of having the expected $e^{-2\varphi}$
dependence of tree-level terms\ref\Wittdim{E. Witten, Nucl. Phys. B443
(1995) 85.}.

Recently, a new formalism for the superstring has been discovered with
local N=2 worldsheet superconformal invariance\ref\meold{
N. Berkovits, Nucl. Phys. B395 (1993) 77.}. This formalism is
related to the RNS formalism by a field-redefinition\ref\medef{
N. Berkovits, Nucl. Phys. B420 (1994) 332.}, but has
the advantage of being manifestly spacetime supersymmetric.
It is especially
well-suited for compactifications to four dimensions, where it allows
manifestly $4D$ super-Poincar\'e invariant quantization\ref\mefour{
N. Berkovits, Nucl. Phys. B431 (1994) 258.}.

In this paper, we use this new formalism to define a two-dimensional
sigma model and construct a superspace effective action for $4D$
compactifications of heterotic and Type II superstrings. While
the heterotic
effective
action in N=1 $4D$ superspace
has already appeared in the literature\ref\cec{S. Cecotti,
S. Ferrara and M. Villasante, Int. J. Mod. Phys. A2 (1987) 1839\semi
J.-P. Derendinger, F. Quevedo and M. Quir\'os,
Nucl. Phys. B428 (1994) 282.}, the Type II
effective action in N=2 $4D$ superspace is new. Since
spacetime-supersymmetry is manifest, there is no distinction between
Ramond and Neveu-Schwarz fields and the previous confusion over the
Ramond sector is easily resolved.
Furthermore, the requirement that the Fradkin-Tseytlin term for
dilaton coupling contains
N=2 worldsheet supersymmetry
implies certain non-renormalization theorems for the effective action.
For those readers who are only interested in the new Type II superspace
effective action and the resulting non-renormalization theorems,
it may be useful to just read section 3 and section 5, and skip
the derivation from the sigma model.

In section 2, we review the new manifestly spacetime supersymmetric
formalism for the superstring. This formalism has critical N=2 worldsheet
superconformal invariance, and for compactifications to four dimensions,
superspace chirality is related to worldsheet chirality\mefour. Superstring
scattering
amplitudes can be computed by calculating correlation functions of
BRST-invariant vertex
operators on N=2 super-Riemann surfaces\meold, and the amplitudes
agree with those obtained using the RNS formalism.
Since the massless vertex
operators are related to the linearized interactions of the sigma model,
these vertex operators will be reviewed.

The massless vertex operators
are manifestly spacetime supersymmetric
and are constructed from prepotentials of the spacetime superfields.
For the heterotic superstring, these prepotentials describe N=1 conformal
supergravity, a tensor multiplet, super-Yang-Mills multiplets, and
chiral scalar multiplets which come from compactification moduli. (We assume
throughout this paper that the compactification manifold has no isometries,
so all moduli are described by four-dimensional scalars.)
For the Type II superstring, the prepotentials describe N=2 conformal
supergravity, a tensor hypermultiplet, and
chiral or twisted-chiral multiplets from the compactification moduli.
The superspace chirality of the compactification
moduli superfields comes from the worldsheet
chirality or twisted-chirality of the relevant N=(2,2) primary fields.
As will be discussed, Type II chiral multiplets can be interpreted as
vector multiplets\ref\vec{R. Grimm, M. Sohnius and J. Wess,
Nucl. Phys. B133 (1978) 275.} while
twisted-chiral multiplets can be interpreted
as tensor hypermultiplets\ref\ten{J. Wess, Acta Phys. Aust. 41
(1975) 409.}. Note that tensor hypermultiplets contain the
same on-shell component fields as scalar hypermultiplets, which are more
commonly used to describe Type II compactification moduli.

Unlike the massless
vertex operators, sigma models and effective actions are constructed
from superspace gauge fields and field strengths
rather than prepotentials. In section 3, we will
review the torsion constraints which relate these gauge fields and
field strengths to their prepotentials.
To analyze these torsion constraints and to facilitate the
construction of
superspace actions,
it will be useful to introduce conformal compensators. (In the
bosonic string, the spacetime dilaton which couples to worldsheet curvature
plays the role of a conformal compensator\ref\siegdil{
W. Siegel, Phys. Lett. 211B (1988) 55\semi
W. Siegel, Phys. Rev. D47 (1993) 5453\semi
W. Siegel, Phys. Rev. D48 (1993) 2826.}.)
Although there are
various types of compensators one can introduce for N=1
and N=2 supergravity, worldsheet
supersymmetry of the Fradkin-Tseytlin term in the
sigma model uniquely determines the correct type. As discussed in reference
\ref\siegalt{W. Siegel, ``Curved extended superspace from Yang-Mills
theory \`a la strings'', preprint ITP-SB-95-42.},
the correct type of compensator can also be determined by requiring
factorization of closed superstring
states involving worldsheet ghosts.

For the heterotic superstring, the compensator must be a chiral scalar
multiplet,
which
identifies the off-shell theory as matter coupled to old minimal supergravity.
For the
Type II superstring, two compensators are required, one which is chiral
(a vector multiplet) and one which is twisted-chiral
(a tensor hypermultiplet).
Normally the compensators are gauge-fixed to remove the unphysical spacetime
invariances. However, in the superstring sigma model,
spacetime conformal invariance is instead removed
by gauge-fixing the non-compensator tensor multiplet. (This is similar
to the bosonic string sigma model, where the physical scalar
is gauge-fixed instead of the dilaton.) For the heterotic
superstring, this leaves a spacetime U(1) invariance (which is related to
worldsheet U(1) invariance), while for the Type II superstring, this
leaves a spacetime U(1)$\times$U(1) invariance (which is related to
worldsheet U(1)$\times$U(1) invariance).

In section 4, we use the above spacetime superfields to
explicitly construct a sigma
model for four-dimensional compactifications of heterotic and Type
II superstrings.
The sigma model for the heterotic superstring was
constructed with the assistance of Jan de Boer, Peter van Nieuwenhuizen,
Martin Ro\v cek, Ergin Sezgin, Kostas
Skenderis, and Kellogg Stelle.
Like the standard $4D$
GS sigma model\ref\GSfour{E. Bergshoeff, E. Sezgin and P. Townsend,
Phys. Lett. 169B (1986) 191\semi
S. J. Gates, Jr., P. Majumdar, R. Oerter and A. van de Ven,
Phys. Lett. 214B (1988) 26.}, this sigma model
contains
a term proportional to $1/\a '$ where the superfields couple
to their massless vertex operators.
In the presence of torsion constraints (which do not put the superfields
on-shell),
this $1/\a '$ term is invariant under classical worldsheet superconformal
transformations.

However, unlike the standard GS sigma model, this sigma model also contains
a Fradkin-Tseytlin term where the spacetime compensators couple to
worldsheet supercurvature.
For the heterotic superstring,
the spacetime chiral compensator couples
to N=(2,0) supercurvature, which is described by a worldsheet chiral
superfield\ref\mesigma{N. Berkovits, Phys. Lett. 304B (1993) 249}.
For the Type II superstring, the spacetime chiral and
twisted-chiral compensators couple to N=(2,2) supercurvature, which is
described by a worldsheet chiral and twisted-chiral superfield.
Quantum
N=2 superconformal invariance of the combined $1/\a '$ and
Fradkin-Tseytlin terms is expected to
imply the equations of motion for the spacetime superfields.
This is currently being checked for the heterotic superstring by
de Boer and Skenderis.

Finally in section 5, we use some simple properties of the sigma model to
construct superspace effective actions. (For
those readers only interested in effective actions, they can skip directly
to this section, using section 3 as a reference.)
Although the heterotic superspace effective action has already appeared
in the literature\cec,
the Type II superspace effective action is new. This N=2 $4D$ superspace
action includes a chiral term for vector multiplet interactions and
a twisted-chiral term for tensor hypermultiplet interactions. The
twisted-chiral term is made SU(2) invariant by introducing
harmonic-like variables\ref\harm{A. Galperin,
E. Ivanov, S. Kalitzin, V. Ogievetsky and E. Sokatchev,
Class. Quant. Grav. 1 (1984) 469.}\ref\roc{A. Karlhede, U. Lindstrom and
M. Ro\v cek, Phys. Lett. 147B (1984) 297\semi
W. Siegel, Phys. Lett. 153B (1985) 51.}.
Since worldsheet
Euler number couples in the Type II sigma model
to the sum of the vector and
tensor compensators, it is straightforward to determine the string-loop
order of each term in the effective action. This type of dilaton coupling
contradicts the standard folklore that loops are counted by
just a hypermultiplet\ref\second{A. Strominger,
``Massless black holes and conifolds in string theory'', hep-th
9504090\semi S. Ferrara, J. Harvey, A. Strominger
and C. Vafa, ``Second-quantized mirror symmetry'', hep-th 9505162.},
and explains the dilaton coupling of
Ramond-Ramond fields. (By dilaton, we always mean the field
which couples to
worldsheet curvature, and not the physical scalar which couples like
the trace of the metric. The confusion in the literature was caused by
the fact that
the physical scalar, which does not count string loops, sits in
a hypermultiplet.)

We then prove various non-renormalization theorems for the
Type II effective action, including the
theorem that the chiral term for vector multiplet
interactions receives
no quantum perturbative or non-perturbative corrections. Since
mirror symmetry of the sigma model
relates chiral and twisted-chiral terms in the
effective action, hypermultiplet
interactions can only receive quantum corrections if mirror symmetry were
broken. This would seem to contradict
Type II/heterotic string-duality conjectures, which require
that Type II hypermultiplet interactions receive quantum corrections\second.
However,
the non-renormalization of
hypermultiplet interactions is related to a
Pecci-Quinn-like symmetry. If this Pecci-Quinn-like
symmetry were non-perturbatively
broken by spacetime instantons
(which would imply the non-perturbative breaking of mirror
symmetry), hypermultiplet interactions could receive quantum corrections.

In section 6, we summarize our results and discuss possible generalizations
of this work for the superstring in more than four dimensions.

\newsec { Review of the New Superstring Description}

By twisting the ghost sector, any critical N=1 string can be
``embedded'' in a critical N=2 string without changing the
physical theory\ref\mevafa{N. Berkovits and C. Vafa,
Mod. Phys. Lett. A9 (1994) 653.}.
 After performing this embedding for the critical
RNS superstring, a field-redefinition allows the resulting N=2
string to be made manifestly spacetime supersymmetric. This N=2
description of the superstring is especially elegant for compactifications
to four dimensions, where the critical $c=6$ matter sector splits into
a $c=-3$ four-dimensional part and a $c=9$ compactification-dependent
part\mefour.

The four-dimensional part of the matter sector contains
the spacetime variables, $x^m$ ($m=0$ to 3), the left-moving
fermionic
variables, $\t^\a$ and $\tb^\ad$ ($\a,\ad=1$ to 2), the conjugate
left-moving
fermionic variables, $p_\a$ and $\bar p_\ad$, and one left-moving
boson $\rho$ (which takes values on a circle of radius 1).
For the heterotic GS superstring, one also has the
right-moving
fermions, $\zeta_q$ ($q=1$ to $32-2r$), which describe
the unbroken gauge degrees of freedom (e.g. the gauge group
is $E_6\times E_8$ when $r=3$, and the gauge group is SO(10)$\times E_8$
when $r=4$).
For the Type II superstring, one needs
the right-moving fermionic fields,
$\th^\ah$,$\tbh{}^\adh$,
$\hat p_\ah$, $\hat{ \bar p}_\adh$, and one right-moving
boson $\hat\rho$.

The compactification-dependent part of the matter sector is described
as in the RNS formalism by a
$c=(9,6+r)$ N=(2,0) superconformal field theory for the
heterotic superstring, and a $c=(9,9)$ N=(2,2) superconformal field theory
for the Type II superstring. Note that because of manifest spacetime
supersymmetry, there is no need to perform a GSO projection
in either the four-dimensional or compactification-dependent sector.

\subsec {The worldsheet action and N=2 stress-energy tensor}

In N=2 superconformal gauge, the worldsheet action for these fields is:
\eqn\aaa{Heterotic: \quad{1\over{\a '}}
\int dz^+ dz^- [\half\dzp x^m \dzm x_m + p_\a \dzp\t^\a +
\bar p_\ad \dzp\tb^\ad +\zeta_q\dzm\zeta_q}
$$ -
{{\a '}\over 2}\dzp \rho D_-\rho + S_C]
$$
\eqn\aab{Type~II:\quad
{1\over{\a '}}
\int dz^+ dz^- [\half\dzp x^m \dzm x_m + p_\a \dzp\t^\a +
\bar p_\ad \dzp\tb^\ad -{{\a '}\over 2}\dzp \rho D_-\rho +}
$$
\hat p_\ah \dzm\th^\ah +
\hat{\bar p}_\adh
\dzm\tbh{}^\adh -{{\a '}\over 2}\dzm \hat\rho D_+\hat\rho
+ S_C ]$$
where  $S_C$ is the action for the compactification-dependent superconformal
field theory, $D_-\rho=\p_-\rho+a_-$ and
$ D_+\hat\rho=\p_+\hat\rho+\hat a_+$ are the
worldsheet covariant derivatives ($e^\rho$ will carry
U(1) charge), and $a_\pm$, $\hat a_\pm$ are the
worldsheet U(1) gauge fields which in superconformal gauge satisfy
$a_+=\hat a_-=0$. (For the Type II superstring, we use the U(1)$\times$U(1)
form of N=(2,2) supergravity which contains two independent U(1) gauge
fields\ref\howegris
{P. Howe and G. Papodopoulos, Class. Quant. Grav. 4 (1987)
11\semi M. Grisaru and M. Wehlau,
Int. J. Mod. Phys. A10 (1995) 753\semi
M. Grisaru and M. Wehlau, ``Superspace measures, invariant actions,
and component projection formulae for (2,2) supergravity'', hep-th
9508139.}.) Note that the equations of motion for $a_-$ and $\hat a_+$ imply
the chirality conditions for $\rho$ and $\hat\rho$. (We are ignoring
subtleties associated with the propagation of $a_-$ and $\hat a_+$.)

The free-field OPE's for these worldsheet variables are
\eqn\azz{x^m(y) x^n(z)\to -\a '\eta^{mn}\log|y-z|,
\quad \rho(y) \rho(z) \to \log(y^- -z^-),}
$$p_\a(y)\theta^\b (z)\to {\a '\delta_\a^\b\over{y^- -z^-}},\quad
\bar p_\ad(y)\bar\theta^\bd (z)\to {\a '\delta_\ad^\bd\over{y^- -z^-}},\quad
\zeta_q(y)\zeta_r(z)\to {\a '\delta_{qr}\over{y^+ -z^+}},$$
$$ \hat p_\a(y)\hat\theta^\b (z)\to {{\a '\delta_\a^\b}\over{y^+ -z^+}},\quad
\hat{\bar p}_\ad(y)\hat{\bar\theta}^\bd (z)\to
{{\a '\delta_\ad^\bd}\over{y^+ -z^+}},\quad
\hat\rho(y) \hat\rho(z) \to \log(y^+ -z^+)$$
where $\eta^{mn}=(-1,1,1,1)$.
Note that the chiral boson $\rho$ has a timelike signature and can not
be fermionized since
$e^{i\rho(y)}~e^{i\rho(z)}~\to e^{2i\rho(z)}/(y^- -z^-)$ while
$e^{i\rho(y)}~e^{-i\rho(z)}~\to (y^- -z^-)$. It has the same behavior as the
negative-energy field $\phi$ that appears when bosonizing the RNS ghosts
$\gamma=\eta e^{i\phi}$ and $\beta=\partial\xi e^{-i\phi}$.
The strange $\a '$ dependence of $\rho$ in \aaa
will later be shown to be related
to the Fradkin-Tseytlin term.

The left-moving $c=6$ stress-energy tensor for this N=2 string is:
\eqn\aac{L=\half\dzm x^m \dzm x_m +
p_\a\dzm \t^\a + \bar p_\ad \dzm\tb^\ad -{\a '\over 2}
\dzm\rho\dzm\rho ~+L_C,}
$$G={1\over \sqrt{\a '}}e^{i\rho} (d)^2 ~+G_C, \quad
\bar G={1\over \sqrt{\a '}}e^{-i\rho} (\bar d)^2 ~+\bar G_C, \quad
J=i\a '\dzm\rho~+J_C,
$$
where
\eqn\aad{d_\a=p_\a+i\s^m_{\a\ad}\tba\dzm x_m -\half(\tb)^2\dzm\t_\a
+{1\over 4}\t_\a \dzm (\tb)^2}
$$
 \bar d_\ad=\bar p_\ad
+i\s^m_{\a\ad}\ta\dzm x_m -\half(\t)^2\dzm\tb_\ad
+{1\over 4}\tb_\ad \dzm (\t)^2,
$$
$(d)^2$ means
$\half\epsilon^{\a\b} d_\a d_\b$, and $[L_C,G_C,\bar G_C,J_C]$
are the left-moving generators of the $c=9$ N=2
compactification-dependent stress-energy tensor.
As was shown in reference \ref\siegGS{W. Siegel, Nucl. Phys. B263 (1986) 93.},
$d_\a$ and $\bar d_\ad$ satisfy
the OPE that $d_\a(y)$ $d_\b(z)$ is regular,
$d^\a (y) \bar d^\ad(z) \to 2i\a '\s_m^{\a\ad}\Pi^m_-/(y^- -z^-)$ where
$\Pi^m_\pm=\p_\pm x^m -i\sigma^m_{\a\ad}(
\t^\a\p_\pm\tb^\ad+\tb^\ad\p_\pm\t^\a),$
and $d_\a(y) \Pi_-^m(z) \to -2i\a '\sigma^m_{\a\ad} \dzm\tb^\ad/(y^- -z^-)$.

For the heterotic superstring, the right-moving $c=26$ N=0
stress-energy tensor is:
\eqn\aae{Heterotic:
\quad\hat L=\half\dzp x^m \dzp x_m + \zeta_q \dzp\zeta_q~+\hat L_C}
where $\hat L_C$ is the right-moving $c=6+r$ compactification-dependent
stress-energy tensor.
For the Type IIB superstring, the right-moving $c=6$
N=2 stress-energy tensor is
\eqn\aaf{Type IIB:\quad\hat L=\half\dzp x^m \dzp x_m +
\hat p_\ah\dzp \hat\t^\ah + \hat{\bar p}_\adh \dzp\tbh{}_\adh -
{\a '\over 2}\dzp\hat
\rho\dzp\hat\rho ~+\hat L_C,}
$$\hat G={1\over\sqrt{\a '}}e^{i\hat\rho} (\hat d)^2 ~+\hat G_C, \quad
\hat{\bar G}={1\over\sqrt{\a '}}
e^{-i\hat\rho} (\hat{\bar d})^2 ~+\hat{\bar G}_C, \quad
\hat J=i\a '\dzp\hat\rho~+\hat J_C,
$$
where $\hat d_\ah$
and $\hat{\bar d}_\adh$
are
obtained from \aad by using hatted variables and replacing
$\dzm$ with $\dzp$.
Since the mirror transformation\ref\mir{S. Hosono, A. Klemm and
S. Theisen, ``Lectures on mirror symmetry'', hep-th 9403096.}
flips the sign
of $\hat J_C$ and switches
$\bar G_C$ with $\hat{\bar G}_C$, the right-moving stress-energy tensor
for Type IIA compactifications is
\eqn\aah{Type IIA:\quad\hat L=\half\dzp x^m \dzp x_m +
\hat p_\ah\dzp \th^\ah + \hat{\bar p}_\adh \dzp\tbh{}_\adh -{\a '\over 2}
\dzp\hat
\rho\dzp\hat\rho ~+\hat L_C,}
$$\hat G={1\over \sqrt{\a '}}e^{i\hat\rho} (\hat d)^2 ~+\hat {\bar G}_C, \quad
\hat{\bar G}=
{1\over \sqrt{\a '}}e^{-i\hat\rho} (\hat{\bar d})^2 ~+\hat G_C, \quad
\hat J=i\a '\dzp\hat\rho~-\hat J_C.$$

The advantage of working with the variables $d_\a$ and $\Pi^m$ is that they
commute with the spacetime supersymmetry generators,
\eqn\aai{q_\a
=\int dz^- [p_\a -i\s^m_{\a\ad}\tba\dzm x_m-{1\over 4}(\tb)^2\dzm\t_\a],}
$$
 \bar q_\ad=\int dz^- [\bar p_\ad
-i\s^m_{\a\ad}\ta\dzm x_m-{1\over 4}(\t)^2\dzm\tb_\ad].
$$
$$\hat q_\ah=\int dz^+ [\hat p_\ah -i\s^m_{\ah\adh}{\tbh}{}^\adh\dzp x_m
-{1\over 4}(\tbh)^2\dzp\th_\ah],$$
$$\hat{\bar q}_\adh=\int dz^+ [\hat{\bar p}_\adh
-i\s^m_{\ah\adh}\th^\ah\dzp x_m-{1\over 4}(\th)^2\dzp\tbh_\adh].
$$
When written in terms of the supersymmetric variables, the actions of
\aaa and \aab
take the more familiar forms
\eqn\aaj{Heterotic:\quad
{1\over{\a '}}
\int dz^+ dz^- [\half\Pi^m_+ \Pi_{m-} + \Pi_-^m T_{m+}-\Pi_+^m T_{m-}+
 \zeta_q\dzm\zeta_q }
 $$+ d_\a \dzp\t^\a + \bar d_\ad \dzp\tb^\ad -
{{\a '}\over 2}\dzp \rho D_-\rho + S_C]$$

\eqn\aak{Type~II:\quad
{1\over{\a '}}
\int dz^+ dz^- [\half\Pi^m_+ \Pi_{m-} + \Pi_-^m (T_{m+}+\hat T_{m+})
- \Pi_+^m (T_{m-}+\hat T_{m-}) }
$$+T_+^m \hat T_{m-}-T_-^m \hat T_{m+}$$
$$
+d_\a \dzp\t^\a +
\bar d_\ad \dzp\tb^\ad +
\hat d_\ah \dzm\th^\ah +
\hat{\bar d_\adh} \dzm\tbh{}^\adh +S_C-
{{\a '}\over 2}(\dzp \rho D_-\rho
+\dzm \hat\rho D_+\hat\rho)]
$$
where $T_{m\pm}=
\s_m^{\a\ad}(\t_\a\p_\pm\tb_\ad
+\tb_\ad\p_\pm\t_\a)$ and $\hat T_{m\pm}=
\s_m^{\ah\adh}(\th_\ah\p_\pm\tbh_\adh
+\tbh_\adh\p_\pm\th_\ah)$.
For the Type II superstring,
$\Pi^m_\pm=\p_\pm x^m +i\s^m_{\a\ad}
(\t^\ad\p_\pm\t^\a +\t^\a\p_\pm\tb^\ad+
\th^\adh\p_\pm\th^\ah +\th^\ah\p_\pm\tbh{}^\adh)$ and $d_\a$ differs from
\aad
by terms which vanish on-shell. If the $d_\a$ and $\rho$ contributions are
dropped, the actions in \aaj and \aak are the standard heterotic and Type II
Green-Schwarz four-dimensional actions\GSfour.

Although one can formally write an N=2 worldsheet supersymmetric version
of these actions as
\eqn\aam{S=\int dz^+ dz^- {1\over {\a ' \det e}}[e_{+-} I + e_{++} L +
e_{--} \hat L +\xi_+ G+\bar\xi_+\bar G +a_+ J}
$$(
+\hat\xi_- \hat G+{\hat{\bar\xi}}_-\hat{\bar G}
+\hat a_- \hat J)]$$
where $(e_{\pm\pm},\xi_\pm,\bar\xi_\pm,a_\pm; \hat\xi_\pm,
\hat{\bar\xi}_\pm,\hat a_\pm)$ are the worldsheet supergravity
fields, the quantum behavior of the $\rho$ field makes it difficult to
make manifest the worldsheet supersymmetry.
Nevertheless, the free-field OPE's of \azz make it straightforward to
check that $[L,G,\bar G, J]$ form a $c=6$ N=2 superconformal
algebra, thereby implying the quantum N=2 superconformal invariance
of \aam.

Note that for the heterotic superstring, if a spacetime superfield $\P$
is worldsheet chiral (i.e. $\bar G$ has no singularities with $\P$), it
is automatically
superspace chiral (i.e. $\Nb_\ad \P=0$
where $\Nb_\ad={\partial\over{\partial
\tb^\ad}}+i\s^m_{\a\ad}\t^\a\p_m$)
since $\bar d_\ad$ has no
poles with $\P$. Similarly for the the Type II superstring, if a spacetime N=2
superfield $\P$ is worldsheet chiral (i.e.
$\bar G$ and $\hat{\bar G}$ have no singularities
with $\P$) or worldsheet twisted-chiral
(i.e. $\bar G$ and $\hat G$ have no singularities
with $\P$), then it is automatically superspace chiral
( i.e., $\Nb_\ad \P=\Nbh_\adh \P=0$) or superspace twisted-chiral
(i.e. $\Nb_\ad \P=\Nh_\ah \P=0$).
Also note that
$\hat G_C$ goes with $\hat G$ for
Type IIB compactifications, while
$\hat{\bar G}_C$ goes with $\hat G$ for Type IIA compactifications.
So for Type IIB (or Type IIA) compactifications, right-moving
worldsheet chirality is correlated (or anti-correlated) with
right-moving chirality of the
compactification-dependent $c=9$ superconformal
field theory.

Scattering amplitudes can be calculated by evaluating correlation
functions of physical vertex operators on N=2 super-Riemann surfaces.
All vertex operators of zero ghost-number are constructed from U(1)-neutral
combinations of the worldsheet matter fields and must carry integer
$J_C$ charge in order not to have branch cuts with $G$.
The massless vertex operators are simpler than in the RNS formalism, and
since they
play an essential role in the construction of the two-dimensional sigma
model, they will be reviewed in the following sub-section.

\subsec { Massless vertex operators }

For the heterotic superstring, all massless
vertex operators which are independent of the
compactification are constructed from the real spacetime superfields
$V_I(x,\t,\tb)$
and $V_m(x,\t,\tb)$, where $V_I$ is the prepotential for super Yang-Mills
($I$=1 to $d$
labels the group index) and $V_m$ is the prepotential for N=1 supergravity
plus a tensor multiplet
($m=0$ to 3 is a spacetime vector index).

In integrated form, these vertex operators are given by\mefour
\eqn\aan{
\int dz^+ dz^-  \{\bar G ,[G, V_P]\}O^P}
where $O^I=j^I$ for the super Yang-Mills vertex operator ($j^I$ is
the right-moving current constructed from the $\zeta_q$'s) and
$O^m=\Pi_+^m=\dzp x^m+i\s^m_{\a\ad}(\t^\a\dzp\tb^\ad+
\tb^\ad\dzp\t^\a)$ for the supergravity/tensor vertex operator.
Note that
$[G,V]$ means the residue of the single pole in the OPE of $G$ and $V$
(for $V$ on-shell, there are no double poles).

Up to surface terms, \aan is equal to
\eqn\aao{\int dz^+ dz^- [ \bar d^\ad ~(\N)^2\Nb_\ad -
  d^\a ~\N_\a(\Nb)^2  +\dzm\tb^\ad \Nb_\ad -\dzm\t^\a~\N_\a}
$$-{i\over 2}
\Pi_-^m\s_m^{\a\ad}[\N_\a,\Nb_\ad] ] V_P(x,\t,\tb) O^P.
$$
Gauge transformations which leave this vertex operator invariant are
$$\d V_P=(\N)^2\L_P +(\Nb)^2 \Lb_P +\d_P^m \p_m \Omega,$$ which
can gauge-fix $4d+4d$ component fields in $V_I$ and
20+20 component fields of $V_m$. In Wess-Zumino gauge,
the remaining $4d+4d$ component fields of $V_I$ are described by
\eqn\aap{V_I=A_{Im}\s^m_{\a\ad}\t^\a\tb^\ad+
\psi_{I\a}\t^\a(\tb)^2 +
\bar\psi_{I\ad}(\t)^2\tb^\ad +D_I(\t)^2(\tb)^2}
where $A_{Im}$ are the gluons, $\psi_{I\a}$ are the gluinos, and $D_I$ is the
real auxiliary field.
In Wess-Zumino gauge,
the remaining 12+12 component fields of $V_m$ are described by
\eqn\aaq{V_m=(h_{mn}+b_{mn}+l\eta_{mn})\s^n_{\a\ad} \t^\a\tb^\ad
+ }
$$(\chi_{m\a}+
\bar\xi^\ad\s_{m\a\ad}) \t^\a(\tb)^2
+(\bar\chi_{m\ad}+
\xi^\a\s_{m\a\ad}) (\t)^2\tb^\ad +D_m (\t)^2(\tb)^2.$$
 From the superspin 3/2 piece of $V_m$ representing conformal
supergravity,
$h_{mn}$ is the traceless graviton,
$D_m$ is the
auxiliary U(1)
gauge field,  and $\chi_{m\a}$ is the gravitino ($\s^m_{\a\ad}\chi_m^\a=0$).
 From the superspin 1/2 piece of $V_m$ representing the tensor multiplet,
$b_{mn}$ is the anti-symmetric tensor, $l$ is the physical scalar,
and $\xi^\a$ is the dilatino.

These vertex operators are on-shell when $V_P$ is an N=2
primary field of weight zero, i.e.
$(\N)^2 V_P=(\Nb)^2 V_P=\p_m\p^m V_P=\p^m V_m=0$.
These imply the usual equations of motion and polarization conditions
on the component fields.

The compactification-dependent massless vertex operators for the
heterotic superstring are constructed
from spacetime chiral superfields $M^{(i)}(x+i\t\tb,\t)$ which couple
to worldsheet chiral primaries $\Omega^{(i)}$ of the $c=(9,6+r)$
N=(2,0) superconformal
field theory representing the compactification manifold ($(i)$ labels
the compactification moduli).
Assuming the compactification has no isometries, the relevant $\Omega^{(i)}$'s
have worldsheet
U(1) charge $+1$ and describe either scalars, vectors, or spinors
of SO($16-2r$).

For scalars, $\Omega^{(i)}$ has
dimension $(\half,1)$ and the vertex operator is
\eqn\aba{\int dz^+ dz^- [\{G, M^{(i)} \Omega^{(i)}\} +
\{\bar G, \bar M^{(i)} \bar\Omega^{(i)}\} ]}
$$=
\int dz^+ dz^- [{1\over \sqrt{\a '}}
e^{i\rho}d^\a (\N_\a M^{(i)})\Omega^{(i)} +M^{(i)}
\{G_C,\Omega^{(i)}\} + c.c. ]$$
where  $c.c.$ means complex conjugate
and we are ignoring double poles in the OPE with $G$ (these double
poles vanish on-shell). For vectors, $\Omega^{(i)}$
has dimension $(\half,\half)$
and the vertex operator is
\eqn\abb{\int dz^+ dz^- [\{G, M^{(i)}_q \Omega^{(i)}\}+
\{\bar G, \bar M^{(i)}_q \bar\Omega^{(i)}\}]\zeta_q}
$$=
\int dz^+ dz^- [{1\over \sqrt{\a '}}
e^{i\rho}d^\a (\N_\a M^{(i)}_q)\Omega^{(i)}
+M^{(i)}_q \{G_C,\Omega^{(i)}\} + c.c. ]
\zeta_q$$
where $q=1$ to $16-2r$.
For spinors,
$\Omega^{(i)}$ has dimension $(\half,{r\over 8})$ and the vertex operator is
\eqn\abc{\int dz^+ dz^- [\{G, M^{(i)}_\gamma \Omega^{(i)}\}
+\{\bar G, \bar M^{(i)}_\gamma\bar\Omega^{(i)}\} ] s^\gamma}
$$=
\int dz^+ dz^- [{1\over \sqrt{\a '}}
e^{i\rho}d^\a (\N_\a M^{(i)}_\g)\Omega^{(i)} +
M^{(i)}_\g \{G_C,\Omega^{(i)}\} + c.c. ]
s^\gamma$$
where $s^\gamma=\exp(\sum_{q=1}^{8-r}\int^z \zeta_{2q-1}\zeta_{2q})$
is a dimension ${8-r}\over 8$ spinor of SO($16-2r$).
As was shown in reference \ref\dist{J. Distler
and B. Greene, Nucl. Phys. B304 (1988) 1.},
these SO($16-2r$) representations combine
into representations of the maximal unbroken subgroup of $E_8$ (e.g.
for $r=3$, they combine into representations of $E_6$).

The compactification-dependent
vertex operators are on-shell when $M^{(i)}$ is primary, i.e.
$(\N)^2 M^{(i)}=0$. In components, this implies
$\p_m \p^m a^{(i)}=\s^m_{\a\ad}\p_m\xi^{\a(i)}=b^{(i)}=0$ where
$M^{(i)}=a^{(i)}(x^+)+\t_\a\xi^{\a(i)}(x^+)+(\t)^2 b^{(i)}(x^+)$ and
$x^+=x+i\t\tb$. (We will supress SO($16-2r$) indices from now on.)

\vskip 20pt

For the Type II superstring, all compactification-independent massless
vertex operators are constructed from a single real superfield,
$U(x,\t,\tb,\th,\tbh)$\ref\riv{V. Rivelles and J.G. Taylor,
J. Phys. A15 (1982) 163.}. In integrated form, these vertex operators
are given by\ref\metop{N. Berkovits and C. Vafa, Nucl. Phys. B433
(1995) 123.}
\eqn\aca{
\int dz^+ dz^- \{\hat{\bar G},[\hat G, \{\bar G ,[G, U]\}]\}=}
$$
\int dz^+ dz^- (\hat {\bar d}{}^\adh ~(\Nh)^2\Nbh_\adh -
 \hat d^\ah ~\Nh_\ah(\Nbh)^2  +\dzm\tbh{}^\adh \Nbh_\adh -\dzm\th^\ah~\Nh_\ah
-{i\over 2}\Pi^m_+ \s_m^{\ah\adh}~[\Nh_\ah,\Nbh_\adh] )$$
$$
( \bar d^\ad ~(\N)^2\Nb_\ad -
  d^\a ~\N_\a(\Nb)^2  +\dzm\tb^\ad \Nb_\ad -\dzm\t^\a~\N_\a
-{i\over 2}\Pi_-^m\s_m^{\a\ad}~[\N_\a,\Nb_\ad]) U$$
where all of the $\N$'s act on $U$.

Gauge transformations which leave this vertex operator invariant are
$$\d U=(\N)^2\L +(\Nb)^2 \Lb +(\Nh)^2\Lh+(\Nbh)^2 \Lbh,$$ which
can gauge-fix 96+96 component fields. In Wess-Zumino gauge,
the remaining 32+32 component fields are described by
\eqn\acd{U=(h_{mn}+b_{mn}+l^{+-} \eta_{mn})\s^m_{\a\ad} \s^n_{\bh\bdh}
\t^\a\tb^\ad
\th^\bh\tbh{}^\bdh}
$$
+(\hat\chi_{m\bh}+\hat{\bar
\xi}{}^\bdh\s_{m\bh\bdh}) \s^m_{\a\ad}\t^\a\tb^\ad\th^\bh(\tbh)^2
+(\chi_{m\a}+
\bar\xi^\ad\s_{m\a\ad}) \s^m_{\bh\bdh}\t^\a(\tb)^2\th^\bh\tbh{}^\bdh
+~c.c.~+$$
$$T_{mn}(\s^{mn}_{\a\bh}\t^\a (\tb)^2 \th^\bh (\tbh)^2
+\bar\s^{mn}_{\ad\bdh}(\t)^2\tb^\ad (\th)^2\tbh{}^\bdh)+$$
$$(A^{++}_m +\p_m l^{++})\s^m_{\a\bdh}\t^\a (\tb)^2 (\th)^2\tbh{}^\bdh+
(A^{--}_m +\p_m l^{--})\s^m_{\ad\bh}(\t)^2 \tb^\ad\th^\bh (\tbh){}^2 +$$
$$y \eps_{\a\bh}\t^\a (\tb)^2 \th^\bh (\tbh)^2
+\bar y \eps_{\ad\bdh}\ (\t)^2 \tb^\ad (\th)^2 \tbh{}^\bdh+$$
$$(A^{U(1)}_m+ A^{+-}_m )\s^m_{\a\ad} \t^\a\tb^\ad (\th)^2 (\tbh)^2 +
(A^{U(1)}_m- A^{+-}_m )\s^m_{\bh\bdh} (\t)^2 (\tb)^2 \th^\bh \tbh{}^\bdh+$$
$$ \psi_\a \t^\a (\tb)^2 (\th)^2 (\tbh)^2 +
\hat\psi_\bh (\t)^2 (\tb)^2 \th^\bh (\tbh)^2 + ~c.c.~+$$
$$D (\t)^2 (\tb)^2 (\th)^2 (\tbh)^2.$$
 From the superspin 1 piece of $U$ representing
the conformal supergravity multiplet,
$h_{mn}$ is the traceless graviton, $\chi_{m\a}$ and $\hat\chi_{m\bh}$
are the gravitinos, $A^{U(1)}_m$ is the U(1) gauge field, $A^{jk}_m$ are the
three SU(2) gauge fields, $T_{mn}$ is the auxiliary tensor, $\psi_\a$
and $\hat\psi_\bh$ are the auxiliary fermions, and $D$ is the
auxiliary scalar. From the superspin 0 piece of $U$
representing the tensor multiplet, $l^{jk}$ is the SU(2)
triplet, $b_{mn}$ is the anti-symmetric tensor, $\xi^\a$ and
$\hat\xi^\bh$ are the dilatinos, and $y$ is a complex auxiliary scalar.

This vertex operator is physical when $U$ satisfies the N=2 primary
conditions
$(\N)^2 U=(\Nb)^2 U=(\Nh)^2 U=(\Nbh)^2 U=\p_m \p^m U=0$, which imply
the usual equations of motion and polarization conditions for the
component fields.

The compactification-dependent massless vertex operators for the
Type II superstring are
constructed from spacetime chiral superfields
$M^{(i)}_c(x+i\t\tb+i\th\tbh,
\t,\th)$ and spacetime twisted-chiral superfields
$M^{(i)}_{tc}(x+i\t\tb-i\th\tbh,
\t,\tbh)$, which couple to the $h^{2,1}$
worldsheet chiral primaries $\Omega_c^{(i)}$ and
the $h^{1,1}$ worldsheet twisted-chiral primaries $\Omega_{tc}^{(i)}$ of the
$c=(9,9)$ N=(2,2) superconformal field theory representing the
compactification manifold ($h^{2,1}$
counts the number of complex moduli and
$h^{1,1}$ counts the number of K\"ahler moduli).
The $\Omega_c^{(i)}$'s have U(1)$\times$U(1)
charge $(+1,+1)$ and dimension $(\half,\half)$, while the
$\Omega_{tc}^{(i)}$'s have U(1)$\times$U(1)
charge $(+1,-1)$ and dimension $(\half,\half)$.

For Type IIB compactifications, the vertex operators are
\eqn\acf{\int dz^+ dz^- (\{\hat G,[G,M_c^{(i)} \Omega_c^{(i)}]\}
+ \{\hat {\bar G},[\bar G,\bar M_c^{(i)} \bar\Omega_c^{(i)}]\})}
$$=
\int dz^+ dz^-
[({1\over\sqrt{\a '}}e^{i\hat\rho}\hat d^\ah\Nh_\ah+\hat G_C)
({1\over\sqrt{\a '}}
e^{i\rho} d^\a\N_\a+ G_C)M_c^{(i)} \Omega_c^{(i)} + c.c.],$$

\eqn\acg{\int dz^+ dz^- (\{\hat {\bar G},[G,M_{tc}^{(i)} \Omega_{tc}^{(i)}]\}
+ \{\hat {G},[\bar G,\bar M_{tc}^{(i)} \bar\Omega_{tc}^{(i)}]\})}
$$=
\int dz^+ dz^- [
({1\over \sqrt{\a '}}
e^{-i\hat\rho}\hat{\bar d}{}^\adh\Nbh_\adh+\hat{\bar G}_C)
({1\over \sqrt{\a '}}
e^{i\rho}d^\a\N_\a+ G_C)M_{tc}^{(i)} \Omega_{tc}^{(i)}+ c.c.]$$
where $\N$'s act only on $M^{(i)}$'s and $G_C$'s act only on
$\Omega^{(i)}$'s.
For Type IIA compactifications, $\Omega_c^{(i)}$ is switched with
$\Omega_{tc}^{(i)}$
and $\hat G_C$ is
switched with $\hat {\bar G}_C$ in the above vertex operators.

These vertex operators are physical when $M_c^{(i)}$ and
$M_{tc}^{(i)}$ satisfy the N=2 primary conditions
$(\N)^2 M_c^{(i)}=
(\Nh)^2 M_c^{(i)}=0$ and $ (\N)^2 M_{tc}^{(i)}=
(\Nbh)^2 M_{tc}^{(i)}=0$.
For constructing superspace actions, it will be useful to separate these
conditions into "reality constraints" (which are satisfied off-shell)
and equations of motion (which are only satisfied on-shell).
The reality constraint for $M_c^{(i)}$ is
$(\N)^2 M_c^{(i)}=(\Nbh)^2 \bar M_c^{(i)}$, which implies the
$8+8$ component expansion of a vector multiplet field strength:
$$M_c^{(i)}=w^{(i)}+\t^\a\xi_\a^{(i)}+\th^\bh\hat\xi_\bh^{(i)}
+
(\t)^2 D_{++}^{(i)}+
\t^\a\th^\bh\eps_{\a\bh}D_{+-}^{(i)}+
(\th)^2 D_{--}^{(i)}+
\t^\a \th^\bh F^{(i)}_{\a\bh} $$
\eqn\aci{+\s^m_{\a\ad}\t^\a(\th)^2\p_m\bar\xi^{\ad (i)}
+\s^m_{\bh\bdh}\th^\bh(\t)^2\p_m\hat{\bar\xi}{}^{\bdh (i)} +
(\t)^2(\th)^2 \p_m
\p^m \bar w^{(i)}}
where $D_{jk}^{(i)}$ is an auxiliary isotriplet, $w^{(i)}$
is a complex scalar,
$F_{\a\bh}^{(i)}$ is a U(1) vector field strength ($\s_m^{\a\ad}\p^m
F_{\a\bh}^{(i)}=\s^m_{\bh\bdh} \p_m F^{\ad\bdh(i)}$),
and $\xi^{(i)}_\a$,
$\hat\xi_\bh^{(i)}$ are SU(2) spinors.

The reality constraint for $M_{tc}^{(i)}$ is
$(\N)^2 M_{tc}^{(i)}=(\Nh)^2 \bar M_{tc}^{(i)}$, which implies the
$8+8$ component expansion of a tensor hypermultiplet field strength:
\eqn\acj{M_{tc}^{(i)}=l_{++}^{(i)}+\t^\a\chi_\a^{(i)}+\tbh{}^\bdh
\hat{\bar\chi}_\bdh^{(i)}+(\t)^2 y^{(i)}+(\tbh)^2 \bar y^{(i)}
+\s^m_{\a\bdh}\t^\a\tbh{}^\bdh (\p_m l_{+-}^{(i)}+
\eps_{mnpq}H^{npq(i)})}
$$
+\s^m_{\a\ad}\t^\a(\tbh)^2\p_m{\bar\chi}^{\ad (i)}
+\s^m_{\bh\bdh}\tbh{}^\bdh(\t)^2\p_m\hat{\chi}^{\bh (i)} +
(\t)^2(\tbh)^2 \p_m
\p^m l_{--}^{(i)}$$
where $l_{jk}^{(i)}$ is a scalar isotriplet, $y^{(i)}$
is a complex auxiliary scalar,
$H_{mnp}^{(i)}$ is a tensor field strength ($\p^m H^{(i)}_{mnp}=0$),
and $\chi^{(i)}_\a$,
$\hat\chi_\bh^{(i)}$ are SU(2) spinors.

The remaining primary conditions,
$(\N)^2 M_c^{(i)}=-(\Nbh)^2 \bar M_c^{(i)}$ and
$(\N)^2 M_{tc}^{(i)}=-(\Nh)^2 \bar M_{tc}^{(i)}$, imply the usual
polarization conditions and equations of motion for these component fields.

\newsec { Superspace Description of the Massless Spectrum}

 From the previous section, we have seen that massless vertex operators
which are independent of the compactification are constructed from
superspace prepotentials. For the heterotic superstring, the N=1
supergravity and tensor multiplets are described by the superspin
$3\over 2$ and $\half$ parts of a real prepotential $V_m$, and the
super-Yang-Mills multiplet is described by a real prepotential $V_I$.
For the Type II superstring, the N=2 supergravity and tensor multiplets
are described by the superspin 1 and 0 parts of a real prepotential $U$.

\subsec { Gauge fields and field strengths}

Although prepotentials are the most compact superspace representations for
these multiplets, they are inconvenient for constructing
super-reparameterization
invariant quantities.
In the sigma model and effective action, it will be more convenient to
represent these multiplets with the vielbein $E_A{}^M$, the
anti-symmetric tensor $B_{MN}$, and the super-Yang-Mills potentials $A^I_M$,
where $A$ labels tangent-superspace indices and $M$ labels curved-superspace
indices. For the heterotic superstring, $A=(a,\a,\ad)$ and $M=(m,\m,\md)$,
while for the Type II superstring, $A=(a,\a j,\ad j)$ and
$M=(m,\m j,\md j)$ where $j=\pm$ are SU(2)-indices which can
be raised and lowered
using the $\eps_{jk}$ tensor. Comparing with the notation from the
previous section,
\eqn\aga{
\t^{\a +}=\t^\a,\quad \t^{\a -}=\th^\a,\quad \tb^{\ad +}=\tbh{}^\ad,\quad
\tb^{\ad -}= -\tb^{\ad}. }
Note that the complex conjugate of $E_M{}^{\a j}$ is
$\epsilon_{jk} E_M{}^{\ad k}$.

The field strengths of these superspace gauge fields are obtained from
the graded commutators
\eqn\agb{[ \N_A,\N_B )= T_{AB}{}^C \N_C +R_{ABC}{}^D M_D{}^C +F_{AB}^I T_I
+f_{AB}^J Y_J,
\quad H_{MNP}=\N_{[M} B_{NP) }, }
where
\eqn\agc{\N_A= E_A{}^M \p_M + A_A^I T_I + w_{AB}{}^C M_C{}^B +\Gamma_A^J Y_J }
is the covariant derivative, $w_{AB}{}^C$ is the
spin connection, $\Gamma_A^J$ are the U(N) connections (N=1 for
heterotic and N=2 for Type II), $M_D{}^C$ are the Lorentz
generators, $T_I$ are the Yang-Mills group generators, $Y_J$
are the U(N) group generators, $T_{AB}{}^C$
is the torsion, $R_{ABC}{}^D$ is the supercurvature, $F^I_{AB}$ is the
super Yang-Mills field strength, $f^J_{AB}$ is the U(N)
field strength, and $H_{MNP}$ is the tensor field strength.
As in ordinary gravity, the connections $w_{AB}{}^C$ and $\Gamma_{A}^J$
are not
independent superfields and are related to $E_A{}^M$ by torsion constraints.

However, since $E_A{}^M$, $B_{MN}$, and $A_M$ contain more component
fields than the prepotentials, one needs to impose further torsion constraints
to remove
the additional degrees of freedom. These constraints will also be needed
for worldsheet supersymmetry of the sigma model and are given
explicitly as constraints 1) and 2) in equations (4.3) and (4.4).
After imposing them, the
above field strengths can be expressed in terms of the
reduced field strengths of the following tables:
$$ \vcenter{\halign{#\hfil&\qquad#\hfil&\qquad#\hfil&\qquad#\hfil
		&\qquad#\hfil\cr
Heterotic multiplet  & tensor & vector & supergravity \cr
\noalign{\vskip4pt\hrule\vskip4pt}
prepotentials & $\Xi_\a$ & $V_I$ & $V_m$ \cr
potentials & $B_{MN}$ & $A^I_M$ & $E_A{}^M$ \cr
unreduced field strengths  & $H_{ABC}$ & $F^I_{AB}$ &
	$T_{AB}{}^C, R_{ABC}{}^D, f_{AB}$  \cr
reduced field strengths  & $L$ & $W^I_\alpha$ &
	$R,G_a, W_{\alpha\beta\gamma}$ \cr}} $$
where the N=1 reduced field strengths are defined by
\eqn\agd{H_{\a\bd c}=\s_{c\a\bd}L,\quad
F_{\ad a}^I=\s_{a \a\ad} W^{I\a},}
$$T_{\a a}{}^\ad=\s_{a\a}{}^\ad R,\quad
T_{\ad a}{}^\ad=G_a,\quad
R_{\ad a}{}^{\b\g}=\s_{a\a\ad} W^{\a\b\g}.$$

$$ \vcenter{\halign{#\hfil&\qquad#\hfil&\qquad#\hfil&\qquad#\hfil\cr
Type II multiplet & tensor & vector & supergravity \cr
\noalign{\vskip4pt\hrule\vskip4pt}
prepotentials & $\Xi$ & $V^{jk}$ & $U$ \cr
potentials & $B_{MN}$ & $A_M$ & $E_A{}^M$ \cr
unreduced field strengths & $H_{ABC}$ & $F_{AB}$ & $T_{AB}{}^C$,
$R_{ABC}{}^D$, $f_{AB}^{jk}$ \cr
reduced field strengths & $L_{jk}$ & $W$ & $S^{jk},G^{jk}_{a},N_{\a\b},
W_{\alpha\beta}$ \cr}}  $$
where the N=2 reduced field strengths are defined by
\eqn\age{H_{\a j\,\ad k\,a}=\s_{a\a\ad} L_{jk},\quad
F_{\ad j\,\bd k}=W \epsilon_{\ad\bd}\eps_{jk},}
$$
T_{a \b j}{}^{\gd k}= i\d_j^k(\s_{a \b\ad}\bar W^{\ad\gd}+\s_a^{\a\gd}
N_{\a\b})
+i \s_{a \b}{}^\gd S^k_j,\quad
T_{a\b j}{}^{\g k}= -2i\s_a^{\g\bd}G^k_{jc}\s^c_{\b\bd}.$$
Note that we have not discussed the chiral prepotential $\Xi$ for
N=1 and N=2 tensor multiplets.
($V_m$ or $U$ can only be used as the tensor prepotential in certain gauges.)
The N=2 vector multiplet
will appear when we discuss compensators and
compactification-dependent states.

Bianchi identities imply various conditions on the above
reduced field strengths.
For example, for the heterotic superstring,
$L$ is a real N=1 linear superfield and $W_\a^I$ is an N=1 chiral
superfield satisfying
\eqn\aha{((\N)^2 +\bar R)L=
((\Nb)^2 + R)L=0,\quad
\Nb_\ad W_\b^I=\N_\a \bar W_\bd^I =\N^\a W_\a^I-\Nb^\ad \bar W^I_\ad=0.}
For the Type II superstring, Bianchi identities imply that
$L_{jk}$ is a real N=2 linear superfield and $W$ is an N=2 chiral
superfield satisfying
\eqn\ahb{\N_{\a (j}L_{kl)}=
\Nb_{\ad (j}L_{kl)}=0,\quad \Nb_{\ad j} W=\N_{\a j}\bar W=
\N^{\a j}\N_{\a k}W-\Nb^{\ad j}\Nb_{\ad k}\bar W=0.}
Note that
$\N_{\a +}
=\N_\a$,
$\Nb_{\ad +}
=\hat\Nb_\adh$,
$\N_{\a -}
=\Nh_\ah$,
$\Nb_{\ad -}
=-\Nb_\ad$, so $L_{--}$ satisfies the twisted-chirality condition
$\Nb_\ad L_{--}=\hat\N_\ah L_{--}=0$.
Implications of Bianchi identities for the reduced N=1 and N=2
supergravity field strengths can be found in references \ref\super
{S.J. Gates, Jr., M. Grisaru, M. Ro\v cek and W. Siegel,
Superspace {\it or}
1001 Lessons in Supersymmetry, Benjamin/Cummings, Reading
(1983).} and \ref\Howe{P. Howe, Nucl. Phys. B199 (1982) 309.}.

\subsec { Compensators}

For constructing superspace actions,
it is useful to introduce compensator superfields which
make the formalism invariant under spacetime scale, U(1), and
for Type II compactifications, SU(2) transformations.
Because of the form of the torsion constraints, it is easy
to generalize a flat superspace action to curved superspace if
the action contains these invariances.
If the original
action does not contain these invariances, one adds an appropriate
power of the compensator superfield so that the trasformation
of the integrand of the original action cancels against the
transformation of the compensator superfield. This makes it
possible to generalize arbitrary flat superspace actions to
curved superspace. Of course, the scale, U(1), and SU(2) invariances are
not physical symmetries (unless they were already present in the
original action), and can be removed by gauge-fixing the compensators
to a constant.

Although it is not widely appreciated, the bosonic string also makes
use of a conformal compensator field, $\varphi$, which couples to
worldsheet curvature and will be called the dilaton\siegdil{}.
Recall that the physical massless vertex operator of the bosonic string
is $\int dz^+ dz^- (g_{mn}+b_{mn}) \p_+ x^m \p_- x^n$, and $g_{mn}$
splits at linearized level
into a traceless piece $h_{mn}=g_{mn}-{1\over D}\eta_{mn}\eta^{pq}
g_{pq}$ representing
conformal supergravity,
and a physical scalar
$l=\eta^{mn}g_{mn}$
which will not be called the dilaton. The vertex operator for
the dilaton, on the other hand, is
constructed from
worldsheet ghosts which, when integrated out, give coupling
to worldsheet curvature.

The usual version of the low-energy effective
action for the bosonic string is
\eqn\aja{  - \int d^D x \, \Phi (4\bo + R + {1\over {12}}H^{abc}H_{abc} )
	\Phi .  }
 (The relative coefficients of these terms can be determined by
T-duality which transforms $g_{mn}$ into $b_{mn}$.
In the supersymmetric cases, they are already fixed by
supersymmetry\siegalt.)  Here $\Phi$ is related to the more common form of the
dilaton field used in string theory by
\eqn\ajb{ \Phi = (-g)^{1/4}e^{-\varphi} }
 (The fact that this is the T-duality invariant combination follows from
the fact that $\Phi$ must soak up the $\sqrt{-g}$ measure, since
$\sqrt{-g}$ is not T-duality invariant.)  At this point we have not yet
seperated out the trace of $g_{mn}$; this we now perform with
full nonlinearity by the Weyl rescaling
\eqn\ajc{ g^{mn} \rightarrow  l^2 g^{mn} }
(We do not scale $\Phi$, and leave $\Phi$-dependence out of the metric
rescaling, so that $\Phi$ stays out of T-duality transformations\siegalt.)
The
result is
\eqn\ajd{ S = - \int d^D x \, \{ \Phi^2 l^2 R + (D-1)\Phi^2(\nabla l)^2}
$$- \left[\Phi^{-1}\nabla(\Phi^2 l)\right]^2
	+ {1\over {12}}\Phi^2 l^6 H^{abc}H_{abc} \} .  $$
 In general relativity one normally breaks the scale invariance
introduced by this rescaling by gauge-fixing $\Phi=(-g)^{1/4}$
($\varphi=0$), or in this case the slightly modified gauge
$\Phi=(-g)^{1/4} l^{-1}$, which produces an action with the standard
form for the  Einstein term, and the right sign for the scalar term:
\eqn\aje{ -\int d^D x \,\sqrt{-g} [ R+ {1\over {12}} l^4 H^2
	+ (D-2)(\nabla\, ln\,l)^2] . }
 However, one can also break scale invariance by choosing the ``string
gauge'' $l=1$, which returns us to the form of the string effective
action before Weyl rescaling.  Although $\Phi$ naively appears to be a
physical scalar in this string gauge, it is easily identified as a
compensating scalar by its ``wrong-sign'' kinetic term. On the other hand,
the physical scalar with a ``right-sign'' kinetic term has turned in string
gauge into the determinant of the metric.

For N=1 and N=2 supergravity, there are various possible types of
conformal compensators\super\ref\dewit{B. de Wit,
R. Phillippe and A. Van Proeyen, Nucl. Phys. B219 (1983) 143\semi
B. de Wit,
P.G. Lauwers and A. Van Proeyen, Nucl. Phys. B255 (1985) 569.}.
However, as will be shown in the following section,
worldsheet supersymmetry of the Fradkin-Tseytlin term uniquely
determines the correct type. (As discussed in \siegalt, the correct compensator
can also be determined by requiring that the closed superstring
dilaton state factorizes into the product of
open superstring ghost states.)

For the heterotic superstring, the conformal compensator is required
to be a superspace chiral and anti-chiral superfield, $\Phi$ and
$\bar\Phi$,
satisfying $\Nb_\ad\Phi=\N_\a\bar\Phi=0$.
Using the U(1) connection $\Gamma_A$ and U(1) generator $Y$,
superspace actions can
be made conformally and U(1) invariant by introducing appropriate powers
of $\Phi$ and $\bar\Phi$. $\Phi$ will be defined to have U(1) weight
$\half$ (i.e. $[Y,\Phi]=\half \Phi$ and
$[Y,\bar\Phi]=-\half\bar\Phi$), so by superspace rules,
it must have conformal weight ${3\over 2}$. (To transform
consistently under superconformal transformations, the conformal weight
of a chiral superfield must be $(4-N)/N$ times its U(1) weight where
$N$ is the number of $4D$ supersymmetries.\super) Note that the
usual choice of U(1) weight
for the chiral compensator is ${1\over 3}$, so we are defining $\Phi$
to be the usual chiral compensator raised to the $3/2$ power. (We
are keeping the convention that $E_\a{}^M$ has conformal and U(1)
weight $\half$.)

For the Type II superstring, worldsheet supersymmetry of the
Fradkin-Tseytlin term requires two different compensators.
One type is described by superspace chiral and anti-chiral superfields,
$\Phi_c$ and $\bar\Phi_c$,
satisfying $\Nb_{\ad}\Phi_c=\Nbh_\ad\Phi_c=\N_{\a }\bar\Phi_c
=\Nh_\a\bar\Phi_c=0$. After imposing
the reality condition $(\N)^2\Phi_c=
(\Nbh)^2\bar\Phi_c$, $\Phi_c$ and $\bar\Phi_c$ can be
identified with $W^{(0)}$ and $\bar W^{(0)}$ where $W^{(0)}$ is the chiral
field strength of a vector multiplet which satisfies equation \ahb. (This
reality condition is not required by worldsheet
supersymmetry of the sigma model, but is necessary
for constructing superspace effective actions.)

The other Type II compensator is
described by superspace twisted-chiral and twisted-anti-chiral superfields,
$\Phi_{tc}$ and $\bar\Phi_{tc}$, satisfying
$\Nb_\ad\Phi_{tc}=\Nh_\ah\Phi_{tc}=
\N_\a\bar\Phi_{tc}=\Nbh_\adh\bar\Phi_{tc}=0$. Although the twisted-chirality
condition on $\Phi_{tc}$ does not look SU(2)-covariant, it can be made
covariant by identifying $\Phi_{tc}$ and $\bar\Phi_{tc}$ with $L^{(0)}_{--}$
and $L^{(0)}_{++}$ where
$L^{(0)}_{jk}$ is the linear field strength of
a tensor hypermultiplet satisfying equation \ahb.
Equation \ahb also implies that $\Phi_{tc}$ satisfies
the reality condition $(\N)^2\Phi_{tc}=(\Nh)^2
\bar\Phi_{tc}$.

To write U(1)$\times$
SU(2)
invariant
superspace actions, one
introduces U(1)$\times$ SU(2) connections
$\Gamma_A^{jk}$,
U(1)$\times$SU(2)
generators
$Y_{jk}$, and appropriate powers of $W^{(0)}$ and $L_{jk}^{(0)}$.
Gauge invariance implies that the component field strength for the
U(1) vector must
carry conformal weight $+2$ and that the component field strength
for the antisymmetric tensor must carry conformal weight $+3$.
Since these field strengths sit in the the $(\t)^2$ components of
$W^{(0)}$ and $L^{(0)}_{jk}$ (see equations \aci and
\acj  for component expansions),
$W^{(0)}$ must carry conformal weight $+1$ and $L^{(0)}_{jk}$ must
carry conformal weight $+2$. Note that $W^{(0)}$ carries U(1) weight
$+1$ and is an SU(2) singlet, while $L^{(0)}_{jk}$ carries U(1) weight zero
and is an SU(2) triplet.

\subsec { Compactification-dependent states}

Finally, we shall give the superfield description of
the compactification dependent massless
states, whose vertex operators are explicitly described
in section 2. Since these moduli superfields describe deformations
of the compactification manifold, we shall assume throughout this
paper that the values of these moduli are small enough so that
the massless spectrum is not modified.

For the heterotic superstring, these states couple to
worldsheet chiral primary fields of the compactification-dependent
$c=9$ N=(2,0)
superconformal field theory. (We shall assume throughout this paper that
the compactification manifold has no isometries, so all moduli are
described by spacetime scalars.) As shown in section 2, worldsheet
and superspace chirality are correlated in the new description of the
superstring, so the massless states coming from compactification
moduli are described by spacetime chiral and anti-chiral
superfields, $M^{(i)}$ and $\bar M^{(i)}$,
satisfying $\Nb_\ad M^{(i)}=\N_\a \bar M^{(i)}=0$. ($(i)$ labels
the compactification moduli, and we are suppressing possible
Yang-Mills group indices.) These moduli superfields will be defined to carry
zero conformal and U(1) weight.

For the Type II superstring, compactification-dependent massless states
couple to the $h^{2,1}$ worldsheet chiral and $h^{1,1}$ worldsheet
twisted-chiral primary fields of the $c=9$ N=(2,2) superconformal
field theory. ($h^{2,1}$ and $h^{1,1}$ count the number of complex
and K\"ahler moduli of the compactification manifold.) As shown in
section 2, left-moving compactification
and superspace chirality are
correlated in the new superstring description, while right-moving
compactification and superspace chirality are correlated (or anti-correlated)
for Type IIB (or Type IIA) compactifications.
Therefore, for Type IIB (or Type IIA) compactifications, the corresponding
massless states are described by $h^{2,1}$ (or $h^{1,1}$)
spacetime chiral and anti-chiral superfields, $M_c^{(i)}$ and
$\bar M_c^{(i)}$,
and by $h^{1,1}$ (or $h^{2,1}$)
spacetime twisted-chiral and
twisted-anti-chiral superfields, $M_{tc}^{(i)}$ and
$\bar M_{tc}^{(i)}$.

As was true for the Type II compensator superfields, it will be useful to
impose a reality condition on these moduli
superfields in order to construct
superspace actions. Although one could use the same reality condition
as for the compensators, this would force the moduli superfields to carry
non-zero U(1) weight. A more convenient choice is to define $M^{(i)}_c$ and
$M_{tc}^{(i)}$ such that
\eqn\ajk{M_c^{(i)}=\log(W^{(i)}/W^{(0)}),\quad M_{tc}^{(i)}=\log(L^{(i)}_{--}/
L^{(0)}_{--})}
where $W^{(i)}$ and $L^{(i)}_{jk}$ satisfy the conditions of equation \ahb
for vector and tensor field strengths. Although this means that $M_c^{(i)}$
and $M_{tc}^{(i)}$ satisfy the non-standard reality conditions,
$(\N)^2 (e^{M_c^{(i)}} W^{(0)})=
(\Nbh)^2 (e^{\bar M_c^{(i)}} \bar W^{(0)})$ and
$(\N)^2 (e^{M_{tc}^{(i)}} L^{(0)}_{++})=
(\Nh)^2 (e^{\bar M_{tc}^{(i)}} L^{(0)}_{--}),$
it allows $M_c^{(i)}$ and $M_{tc}^{(i)}$ to have zero conformal and
U(1) weight.

\newsec { Sigma Models}

Because we know the action in a flat background and the massless
vertex operators from section 2,
it is straightforward to construct a two-dimensional
sigma model for the superstring in a curved background.
This sigma model should be spacetime
super-reparameterization and gauge invariant, and at linearized level,
its interactions should reproduce the massless vertex operators.

When the spacetime superfields satisfy their equations of motion,
the sigma model action is expected to be
N=2 superconformally invariant at the quantum level,
i.e., the components of the
N=2 stress-energy tensor form a $c=6$ N=2 superconformal algebra.
These superspace equations of motion are currently being
computed for the heterotic superstring
by de Boer and Skenderis, which will be the first covariant
$\b$-function computation in a fermionic
background.

\subsec { The classical term}

The two-dimensional sigma model is constructed from the
spacetime superfields of section 3 and splits into a classically
worldsheet N=2 superconformally invariant
term and a Fradkin-Tseytlin term.
Using the action in a flat background and the vertex operators
of section 2, it is
easy to guess the classically worldsheet superconformally invariant
term:

\eqn\ala{
Heterotic:\quad{1\over{\a '}}
\int dz^+ dz^- [\half\Pi^a_+ \Pi_{a-} + B_{AB}\Pi^A_+ \Pi^B_- +
\zeta_q\dzm\zeta_q }
$$+(A_B^I \Pi^B_- +W_\a^I d^\a -\bar W_\ad^I \bar d^\ad)j_I
+
d_\a \Pi^\a_+ +
\bar d_\ad \Pi^\ad_+ +
S_C
+\{G, M^{(i)}\Omega^{(i)}\}
+\{\bar G, \bar M^{(i)}\bar\Omega^{(i)} \}]$$

\eqn\alb{Type IIB:\quad{1\over{\a '}}
\int dz^+ dz^- [\half\Pi^a_+ \Pi_{a-} + B_{AB}\Pi^A_+ \Pi^B_- }
$$+
d_\a \Pi^{\a +}_+ -
\bar d_\ad \Pi^{\ad -}_+
+
\hat d_\ah \Pi^{\ah -}_- +
\hat{\bar d}_\ad \Pi^{\adh +}_- +$$
$$d_\a P^{\a\bh} \hat d_\bh +
\bar d_\ad \bar P^{\ad\bdh} \hat{\bar d}_\bdh +
d_\a Q^{\a\bdh} \hat{\bar d}_\bdh +
\bar d_\ad\bar Q^{\ad\bh} \hat d_\bh  $$
$$
+S_C +
\{G, [\hat G, M_c^{(i)} \Omega^{(i)}_c]\}
+\{\bar G, [\hat{\bar G}, \bar M_{c}^{(i)} \bar\Omega^{(i)}_c]\}$$
$$+
\{G, [\hat{\bar G}, M_{tc}^{(i)} \Omega^{(i)}_{tc}]\}+
\{\bar G, [\hat G, \bar M_{tc}^{(i)} \bar\Omega^{(i)}_{tc}]\} ]$$
where $\Pi^A_\pm$
$=E_M{}^A$
$ \p_\pm Z^M$, $Z^M$
=$(x^m$,
$\t^\mu$,
$\tb^{\dot\mu})$
for
heterotic,
and $Z^M$=$(x^m$,$\t^\mu,$
$\tb^{\dot\mu}$,$\th^\mh$,$\tbh{}^\mdh)$
for Type II.
In a flat background
(i.e., $E_m{}^a=\d^a_m$, $E_\m{}^a=\sigma^a_{\m\md}
\tb^{\md}$,
$B_{\a a}=\s_{a\a\ad}\tb^\ad$ for heterotic;
$E_m{}^a=\d^a_m$, $E_{\m j}{}^a=\sigma^a_{\m\md}
\tb^{\md}_j$,
$B_{\a j a}=\s_{a\a\ad}\tb_j^\ad$,
$B_{\a +\bh -}=\t_\a\th_\bh$,
$B_{\a +\bdh +}=\t_\a\tbh{}_\bdh$ for Type II), it is easy to show that
this action reproduces \aaa and \aab.
Note that we are omitting the kinetic term for
$\rho$ since it is zeroth order in $\a '$, and
therefore goes with the Fradkin-Tseytlin term.

For the Type IIB sigma model,
$M_c^{(i)}$ and $M_{tc}^{(i)}$ are the
chiral and twisted-chiral moduli superfields defined in
section 3. Although the reality constraints on
$M_c^{(i)}$ and $M_{tc}^{(i)}$ are not implied by classical
worldsheet superconformal invariance, these constraints (or a suitable
modification) are expected to
be implied by quantum superconformal invariance.
$P^{\a\b}$ and $Q^{\a\bd}$ are chiral
and twisted-chiral field strengths of N=2 conformal supergravity. From
the
$d^\a \hat d^\ah$ and $d^\a \hat{\bar d}{}^\adh$ terms in the supergravity
vertex operator of \acd, one sees that at linearized level,
$P^{\a\b}=(\Nb)^2 \N^\a (\Nbh)^2 \Nh^\b U$
and $Q^{\a\bd}=(\Nb)^2 \N^\a (\Nh)^2 \Nbh{}^\bd U$.
Note that the Type IIA sigma model action is obtained from the Type IIB
action by switching $\Omega_c^{(i)}$ with $\Omega_{tc}^{(i)}$ and
$\hat G_C$ with $\hat{\bar G}_C$.

Up to field redefinitions (e.g. $E_A{}^M\to E_A{}^M+f_A^B E_B{}^M$
or $d^\a \to d^\a +g_A^\a \Pi_\pm^A$), this
is the most general sigma model action which is
invariant under spacetime super-reparameterizations,
under
the gauge transformations
($\d A^I_B=\N_B \L^I$, $\d W_\a^I=f^I_{JK}\L^J W_\a^K$,
$\d j_I=f^K_{IJ} \L^J j_K$)
and ($\d B_{AB}=\N_A \L_B-\N_B\L_A$), and under classical N=2
worldsheet superconformal transformations.

Classical worldsheet
superconformal transformations are defined by taking the Poisson
bracket with the N=2 stress-energy tensor. (Unfortunately, the
quantum nature of $\rho$ is an obstacle to making N=2
worldsheet supersymmetry manifest.) For example, from the Poisson bracket
with $G$, $\d Z^M=e^{i\rho} d^\a E_\a{}^M$,
$ \d \bar d^\ad=e^{i\rho} \Pi_-^{\a\ad} d_\a$, and
$\d\Pi_\pm^A= e^{i\rho} d^\a T_{\a B}{}^A \Pi^B_{\pm}
+\d_\a^A \p_{\pm}(e^{i\rho} d^\a)$ where $T_{AB}{}^C$ is the torsion
defined in equation \agb.
Under these transformations, \ala and \alb
should be invariant up to terms proportional
to the N=2 stress-energy tensor. (Terms proportional to the stress-energy
tensor can be absorbed by transforming the worldsheet supergravity fields.
We would like to thank Peter van Nieuwenhuizen for
suggesting this method.)
It is straightforward to check that classical invariance
implies that $E_M{}^A$, $B_{AB}$, and $A_M$ satisfy the
following constraints on their field strengths:

\eqn\ama{Heterotic:\quad 1)~ T_{\a\b}{}^c=T_{\a\b}{}^{\dot\gamma}=0,~
F^I_{\a\b}=0;}
$$
2)~T_{\a\bd}{}^c
= \s_{\a\bd}^c,~ T_{\a (b c)}=0,~F^I_{\a\bd}= 0$$
$$3)~ T_{\a c}{}^\bd=0,\quad H_{\a\bd c}=\s_{c\a\bd}.$$

\eqn\amb{Type II:\quad 1)~ T_{\a j\b k}{}^c=T_{\a j\b k}{}^{\dot\gamma}=0;}
$$2)~T_{\a j\bd k}{}^c
= \eps_{jk}\s_{\a\bd}^c, ~T_{\a j (b c)}=0,~F^I_{\a\bd}= 0;$$
$$3)~ T_{\a + \, c}{}^{\bd +}=\s_{c \a\ad} \bar P^{\ad\bd},\quad
T_{\a - \, c}{}^{\bd -}=\s_{c \a\ad} \bar P^{\bd\ad},\quad
T_{\a + \, c}{}^{\bd -}=
T_{\a - \, c}{}^{\bd +}=0,$$
$$
T_{\a +\,  c}{}^{\b -}=\s_{c \a\ad}\bar Q^{\ad\b},\quad
T_{\a -\,  c}{}^{\b +}=\s_{c \a\ad}Q^{\b\ad},\quad
H_{c\a j\bd k }=\s_{c\a\bd }\delta_{jk}.$$

The first type of constraints
are the usual representation preserving constraints
which allow a consistent definition of chiral (and twisted-chiral)
superfields. The second
type are conventional constraints which define the
vector components of the super-vierbein and super
Yang-Mills gauge field, $E_a{}^M$
and $A_a^I$, in terms of the spinor components.
Note that
conventional constraints for the connections $w_{AB}{}^C$,
$\Gamma_A$, and $\Gamma_A^{jk}$ can be defined arbitrarily since
these superfields never appear in the sigma model action.

The third type of constraints are conformal-breaking constraints, which
are necessary since the sigma model action is not invariant under the
spacetime scale transformations that transform
$\d E_a{}^M=\L E_a{}^M$. In their absence for the heterotic superstring,
$T_{\a c}{}^\bd$ and $H_{\a\bd c}$ would satisfy the equations
$T_{\a c}{}^\bd=R\s_{c\a}^\bd$ and $H_{\a\bd c}=L \s_{c\a\bd}$
from equation \agd.
So these constraints break
conformal invariance for the heterotic sigma model
by gauge-fixing $L=1$ (which implies $R=0$ by
the linear condition
$((\Nb)^2 +R)L=0$).

In the absence of conformal-breaking constraints for the
Type II sigma model,
$T_{a \b j}{}^{\gd k}$,
$T_{a \b j}{}^{\g k}$, and $H_{c\a j\bd k}$ would satisfy the conditions
of equation \age \Howe. Therefore conformal and
$SU(2)/U(1)$ invariance is broken in the Type II sigma model
by gauge-fixing $L_{jk}=\delta_{jk}$ (which as will be shown
in the following paragraphs, implies
$N^{\a\b}=G^{a}_{jk}\d^{jk}=
S^{++}=S^{--}=0$).
The conformal-breaking constraint also implies
that $P^{\a\b}=W^{\a\b}+\eps^{\a\b} S^{+-}$ and that
$Q^{\a\bd}= G_{++}^{\a\bd}$.

In the tensor calculus approach to constructing actions for N=2 supergravity
coupled to matter\dewit, a ``minimal multiplet" is employed as a starting
point, consisting of conformal supergravity coupled to a vector multiplet,
in the scale (+U(1)) gauge $W^{(0)}=1$.  Since for string theory it is more
useful to apply the string gauge $L_{jk}=\delta_{jk}$, it is instructive to
compare the field content of the two gauges.
As for N=1\super, both gauges
imply the vanishing of some of the torsions arising from the superconformal
covariant derivatives.
 From \age, the dimension 1 torsions are originally $W_{\alpha\beta}$,
$S_{(jk)}$, $N_{\alpha\beta}$, and $G_j{}^k{}_a$ (hermitian).

The basic procedure is:
(1) Fix {\it one} of the two superfields $W^{(0)}$ or $L_{jk}$. (Although
one could also fix the tensor compensator $L^{(0)}_{jk}$,
this would give the same field content as fixing the non-compensator
$L_{jk}$.)
That breaks scale
invariance (and more), but it doesn't introduce a dimension 0 scalar or
dimension $\half$ spinor (which come from fixing the second compensator),
so you get no new tensors.
(2) Since scale invariance is fixed, the torsions are now tensors.
(Before, they were noncovariant under scale transformations.)  Thus, at
$\theta=0$ they correspond to gauge invariant component fields. By a
component analysis, we know what the component fields are, so match them
up.  The superspace is determined by looking at dimension 1 tensors, and
there is no lower dimension stuff, so a linearized analysis is sufficient.

These are the relevant component fields:
(1) From conformal supergravity, there is the auxiliary field
$W_{\alpha\beta}$, so that torsion will always survive.  There are also the
U(2) {\it gauge} fields $G_j{}^k{}_a$, which will become gauge invariant
only if they find scalars to eat.
(2) A vector multiplet has a gauge vector, with the dimension 1 field
strength $F_{\alpha\beta}$, as well as the hermitian auxiliary fields
$D_{(jk)}$.  It also has a complex physical scalar of dimension 0, whose
real part is eaten by the conformal graviton, and whose imaginary part is
eaten by the U(1) gauge vector.
(3) A tensor multiplet has a dimension 1 field strength $H_a$, and a
complex scalar auxiliary field.  It also has the hermitian physical fields
$L_{(jk)}$ of dimension 0, of which one is eaten by the graviton, the
remaining two by the SU(2)/U(1) gauge vectors.  (These dimensions differ
from the original conformal weights, since scale invariance will be
broken.)

The net result is then, from the original torsions $W_{\alpha\beta}$,
$S_{(jk)}$, $N_{\alpha\beta}$, and $G_j{}^k{}_a$:
(1) For the standard gauge $W^{(0)} = 1$, $W_{\alpha\beta}$ survives as the
conformal supergravity auxiliary, while $N_{\alpha\beta}=F_{\alpha\beta}$.
$S_{(jk)}$ becomes hermitian, identified as the vector multiplet
auxiliaries.  The U(1) part of $G_j{}^k{}_a$ survives, as the U(1) gauge
field that ate the imaginary part of the vector multiplet scalar, but the
SU(2) part does not, since it remains gauge.  The net result is
\eqn\Wc{ W^{(0)} = 1 \quad\Rightarrow\quad
 S_{jk} = \bar S_{jk},\qquad G_{(jk)a} = 0.}
(2) For the string gauge $L_{jk} = \delta_{jk}$, $W_{\alpha\beta}$ again
survives, but there is no field to correspond to $N_{\alpha\beta}$.  Since
the tensor multiplet has only a complex scalar for auxiliaries, with
vanishing weight under the unbroken U(1) subgroup of SU(2), it becomes
$S_{+-}$ (still complex), while $S_{++}$ and $S_{--}$ vanish.  The tensor
field strength $H_a$ replaces the U(1) gauge field in the U(1) part of
$G_j{}^k{}_a$ (as in the analogous N=1 case), while $G_{++a}$ and $G_{--a}$
survive as the SU(2)/U(1) gauge vectors by eating the corresponding
physical scalars of the tensor multiplet.  But $\d^{jk}
G_{jka}$ dies for lack of a
scalar to eat.  The net result for this case is then
\eqn\Lc{
 L_{jk} = \delta_{jk}\quad \Rightarrow\quad
 S_{++}=S_{--}=N_{\alpha\beta}=\d^{jk}G_{jka}=0.}

Actually, the third set of constraints for the heterotic
superstring is slightly inconvenient.
Because of anomalous transformations of the anti-symmetric tensor
field, the linear superfield $L$ is not invariant under Yang-Mills
and Lorentz gauge transformations. The appropriate gauge-invariant
generalization of $L$ is
\eqn\apa{\tilde L = L +\a '(c_1 \Omega_{YM}+c_2 \Omega_{Lorentz})}
where
$\Omega_{YM}$ is the super Yang-Mills Chern-Simons form
satisfying $((\Nb)^2 +R)\Omega_{YM}=W^I_\a W^{I\a}$,
$\Omega_{Lorentz}$ is the super Yang-Mills Chern-Simons form
satisfying $((\Nb)^2 +R)\Omega_{Lorentz}=W_{\a\b\g} W^{\a\b\g}$,
$c_1$ and $c_2$ are compactification-dependent constants, and
$W_{\a\b\g}$ is the chiral field strength of conformal supergravity.

Since $L=1$ breaks Yang-Mills and Lorentz invariance, it is more
convenient to choose the gauge $\tilde L=1$, which implies
$R=c_1 W^I_\a W^{I\a}+c_2 W_{\a\b\g}W^{\a\b\g}$ from the linear
condition on $L$. Therefore, in order to preserve manifest Yang-Mills
and Lorentz invariance, the third type of constraints in \ama
should be
modified for the heterotic superstring to
\eqn\apb{3)~ T_{\a c}{}^\bd=
\a '\s_{c\a}^\bd (c_1 W^I_\g W^{I\g}+c_2 W_{\g\d\kappa}W^{\g\d\kappa}),
{}~\tilde H_{\a\bd c}=\s_{c\a\bd}}
where
\eqn\cer{\tilde H_{MNP}=H_{MNP}}
$$+ \a 'Tr(c_1 (A_{[M}\p_N A_{P)} + {2\over 3} A_{[M} A_N A_{P)})+
+c_2 (w_{[M}\p_N w_{P)}
+{2\over 3} w_{[M} w_N w_{P)}))$$
includes the contribution of the Chern-Simons
forms ($A_M^I$ is the Yang-Mills gauge field and $w_{MB}{}^C$ is the
Lorentz spin connection).

Because these modifications to the heterotic torsion
constraints are higher-order in $\a '$,
they can not be checked using classical worldsheet superconformal invariance.
To justify them,
one should check that with these modified constraints, quantum N=(2,0)
superconformal invariance implies equations of motion which are
invariant under Yang-Mills and Lorentz gauge transformations. Note that,
with the exception of the representation-preserving constraints,
modifications to the torsion constraints can always be absorbed by
compensating transformations on the spacetime superfields. So any
$\a '$ corrections to the non-representation-preserving constraints
can be replaced by $\a '$ corrections to the relations between
spacetime superfields
in the sigma model action.
This is similar to the situation in standard N=2
worldsheet supersymmetric sigma models, where
worldsheet supersymmetry implies
certain relations between coupling constants, but these relations
may receive quantum corrections.

\subsec{ The Fradkin-Tseytlin term}

So the classical term in the sigma model action is \ala or
\alb with the constraints
of \ama (and \apb) or
\amb, and one now needs to construct a Fradkin-Tseytlin term.
As usual, this term is necessary because the massless vertex operator
for the physical scalar, $\int dz^+ dz^- \dzm x^m \dzp x_m$, couples to the
determinant of the spacetime vierbein in the sigma model. So instead of
coupling to this massless vertex operator, the spacetime dilaton couples
to the N=2 supercurvature in a Fradkin-Tseytlin term.

It is often incorrectly stated that the dilaton must sit in the same
supersymmetry multiplet as the anti-symmetric tensor, which leads
to the erroneous conclusion that the dilaton must couple classically.
When spacetime scale invariance is used to gauge-fix
the non-compensator
tensor multiplet (instead of the more conventional gauge-fixing of the
conformal compensator),
the physical anti-symmetric tensor sits in the multiplet of Poincar\'e
supergravity. Although this version of Poincar\'e supergravity does
contain a scalar field, the scalar plays the role of the
determinant of the vierbein, rather than the role of the dilaton.
So the dilaton must sit in the only remaining superfield, which is
the conformal compensator $\Phi=e^{-\phi}$. (Note that $\phi=-\log\Phi$,
rather
than $\Phi$, will appear directly in the sigma model.)

For the heterotic superstring,
the logarithm of the conformal
compensator
couples to N=(2,0) worldsheet supercurvature, which
is described by a worldsheet chiral superfield $\Sigma$
and its complex conjugate $\bar\Sigma$. These worldsheet
superfields are defined by
$[\bar D_{\bar \k},D_+]=\Sigma~ (m+iy)$
and $[D_{\k},D_+]=\bar\Sigma~ (m-iy)$
where $m$ is the Lorentz generator, $y$ is the U(1)
generator,
$D_\k$ and $\bar D_{\bar\k}$ are the covariant fermionic derivatives,
$D_+$ is the covariant right-moving bosonic derivative,
and $\{D_\k,\bar D_{\bar\k}\}=D_-$
where $D_-$ is the covariant left-moving bosonic derivative.
In components, $\Sigma=\chi+\k(r+if)$ and
$\bar\Sigma=\bar\chi+\bar\k(r-if)$ where $\k$ and $\bar\k$ are
the super-worldsheet anti-commuting parameters, $r$ is the two-dimensional
curvature, $f$ is the field strength of the worldsheet U(1) gauge field,
and $\chi$,$\bar\chi$ are the field strengths of the worldsheet
gravitini\ref\gateshet{R. Brooks, S. J. Gates, Jr. and F. Muhammad,
Nucl. Phys. B268 (1986) 599.}.

If and only if $\Phi=e^{-\phi}$
is superspace chiral (and therefore $\{ G,\phi\}=0$),
one can construct the worldsheet supersymmetric Fradkin-Tseytlin term
\eqn\apd{\int dz^+ dz^- (\int d\k \eps^{-1} \phi_\k \Sigma+
 \int d\bar\k {\bar \eps}^{-1}\bar\phi_{\bar\k}\bar\Sigma)=}
$$\int dz^+ dz^- e^{-1}
[(\phi+\bar\phi)r +i(\phi-\bar\phi)f +[G,\phi]\chi+[\bar G,\bar\phi]\bar\chi]
=$$
$$\int dz^+ dz^- e^{-1}
[(\phi+\bar\phi)r +i(\phi-\bar\phi)f +{1\over \sqrt{\a '}
}(e^{i\rho}d^\a\N_\a\phi\chi+
e^{-i\rho}\bar d^\ad \N_\ad\bar\phi\bar\chi)]$$
where $\eps$ is the super-worldsheet chiral density and $e$
is the ordinary worldsheet density,
we ignore possible double
poles in the OPE of $G$ with $V$ (as in the vertex operators of section
2.2),
and $\phi_\k$ is a worldsheet and spacetime superfield  defined such that
$Q_\k\phi_k=[G,\phi_\k]$ and $\bar Q_{\bar\k}\phi_\k=[\bar G,\phi_\k]$  where
$Q_\k$ and $\bar Q_{\bar\k}$ are the worldsheet supersymmetry generators.
(On a flat worldsheet, $Q_\k=\p_\k+\bar\k\p_-$ and
$\bar Q_{\bar\k}=\p_{\bar\k}+\k\p_-$, so
$\phi_\k=\phi+\k[G,\phi]-\k\bar\k \p_-\phi$ satisfies this definition.
It is easily checked that $\phi_\k$ is worldsheet chiral since
$\bar D_{\bar\k}\phi=(\p_{\bar\k}-\k\dzm)\phi=0$.)
Note that the Fradkin-Tseytlin term can be written independently of $\Gamma_A$
since
$\N_\a\phi=E_\a{}^M\p_M\phi+\Gamma_\a=E_\a{}^M \p_M(\phi+\bar\phi)$.

 From the form of the Fradkin-Tseytlin term, it is reasonable to call
the $\t=\tb=0$ component of $\phi+\bar\phi$ the dilaton, and the
$\t=\tb=0$ component of $\phi-\bar\phi$ the axion. Just as the dilaton
zero mode couples to the worldsheet Euler number, the axion zero
mode couples to the worldsheet U(1) instanton number. Note that this axion
is the compensating field for spacetime U(1) transformations, and not the
dual of the anti-symmetric tensor.

When the heterotic Fradkin-Tseytlin term is in
N=(2,0) superconformal gauge, $\phi-\bar\phi$ couples to $a_-$ in
the same way as the $\rho$ field in \aaa, so one can combine these couplings
into $\int dz^+ dz^- \dzp(\rho+i(\phi-\bar\phi))a_-$. This term is
invariant under the spacetime U(1) transformation
\eqn\apg{\d\phi=\L,~\d\bar\phi=-\L,~\d\Gamma_M=\N_M\L,~\d E_\a{}^M=\L
E_\a{}^M,
{}~\d E_\ad{}^M=-\L E_\ad{}^M }
if
one also transforms the worldsheet variables
$$\d \rho=-2i\L,~\d d^\a=\L d^\a,~
{}~\d \bar d^\ad=-\L \bar d^\ad .$$
Although it may look unusual for worldsheet variables to transform under
a spacetime gauge transformation, it is similar to the transformation of
heterotic chiral fermions, $\zeta_q$,
under a Yang-Mills gauge transformation.

However, the kinetic term for the $\rho$ variable, $\int dz^+ dz^-
\dzm\rho\dzp\rho$, is not invariant under the transformation of \apg.
We must therefore modify the kinetic term to
$\int dz^+ dz^-$
$ \dzm
 (\rho+i(\phi-\bar\phi))$
$ \dzp (\rho+i(\phi-\bar\phi))$, which combines with the Fradkin-Tseytlin
term in superconformal gauge
to form the worldsheet and spacetime covariant expression
\eqn\awa{\int dz^+ dz^- [\dzp( \rho+i(\phi-\bar\phi))
D_- (\rho+i(\phi-\bar\phi))}
$$
+(\phi+\bar\phi)r +{1\over\sqrt{\a '}}(e^{i\rho}d^\a\N_\a\phi\chi+
e^{-i\rho}\bar d^\ad \N_\ad\bar\phi\bar\chi)].$$
(Although it may seem strange
to have a term in the sigma model action which is quadratic in $\phi$,
this also occurs in the bosonic string if one couples the dilaton $\varphi$ to
worldsheet ghosts, and then integrates out the ghosts to obtain
$\int dz^+ dz^- (\dzp\varphi\dzm\varphi+\varphi r)$\ref\banks
{T. Banks, D. Nemeschansky and A. Sen, Nucl. Phys. B277 (1986) 67.}.)
Note that the equation of motion for $a_-$ now implies that the
right-moving part of
$\rho$ satisfies $\dzp\rho=-i\dzp(\phi-\bar\phi)$.

\vskip 20pt

For the Type II superstring, the logarithms
of the conformal compensators couple to
N=(2,2) worldsheet supercurvature, which is described by a worldsheet
chiral superfield $\Sigma_c$ and its complex conjugate $\bar\Sigma_c$,
and by a worldsheet twisted-chiral
superfield $\Sigma_{tc}$ and its complex
conjugate $\bar\Sigma_{tc}$. (We use the U(1)$\times$U(1)
form of N=(2,2) supergravity which contains two independent U(1) gauge
fields.\howegris) These worldsheet superfields are defined by
$\{\bar D_{\bar\k},\hat{\bar D}_{\hat{\bar\k}}\}=
\Sigma_c(m+iy)$,
$\{ D_{\k},\hat{ D}_{{\k}}\}=
\bar\Sigma_c(m-iy)$,
$\{\bar D_{\bar\k},\hat{ D}_{\hat{\k}}\}=
\Sigma_{tc}(m+i\hat y)$,
$\{ D_{\k},\hat{\bar D}_{\hat{\bar\k}}\}=
\bar\Sigma_{tc}(m-i\hat y)$,
where $m$ is the Lorentz generator, $y$ and $\hat y$
are the U(1)$\times$U(1) generators, and $D_\k$, $\bar D_{\bar\k}$,
$\hat{D}_{\hat{\k}}$,
$\hat{\bar D}_{\hat{\bar\k}}$ are the covariant fermionic derivatives
satisfying $\{
 D_\k, \bar D_{\bar\k}\}=D_-$ and
$\{ \hat{D}_{\hat{\k}},
\hat{\bar D}_{\hat{\bar\k}}\}=D_+$.

In components, $\Sigma_c=b+\k(\chi+\psi)+\hat\kappa(\hat\chi+\hat\psi)
+\k\hat\k(r+i(f+\hat f)+d)$ and
$\Sigma_{tc}=c+\k(\bar\chi-\bar\psi)+\hat{\bar\kappa}(\hat\chi-\hat\psi)
+\k\hat{\bar\k}(r+i(f-\hat f)-d)$
where $r$ is the two-dimensional curvature, $f$ and $\hat f$ are the right
and left-moving worldsheet U(1) field strengths, $\chi$ and $\hat \chi$
are the complex gravitino field-strenghs, $b$ and $c$ are complex
weight 1 auxiliary fields, $\psi$ and $\hat\psi$ are complex weight
$3\over 2$ auxiliary fields, and $d$ is a real weight 2 auxiliary field.
These additional auxiliary fields are present since we are
using the U(1)$\times$U(1)
formulation of N=(2,2) supergravity.

If, and only if, $\P_c=e^{-\phi_c}$
is superspace chiral and $\P_{tc}=e^{-\phi_{tc}}$ is superspace
twisted-chiral, one can construct the following
Type II worldsheet Fradkin-Tseytlin
term:
\eqn\ara{
\int dz^+ dz^- (\int d\k d\hat\k ~\eps^{-1}_c\phi^c_{\k\hat\k} \Sigma_c +
 \int d\bar\k
d\hat{\bar\k}~ \bar \eps_c^{-1}\bar\phi^c_{\bar\k\hat{\bar\k}}\bar\Sigma_c+}
$$
\int d\k d\hat{\bar\k} ~\eps_{tc}^{-1}\phi^{tc}_{\k\hat{\bar\k}} \Sigma_{tc} +
 \int d\bar\k d\hat{\k}~\bar \eps_{tc}^{-1}
\bar\phi^{tc}_{\bar\k\hat\k}\bar\Sigma_{tc})=$$
$$\int dz^+ dz^-
e^{-1}
[(\phi_c+\bar\phi_c+\phi_{tc}+\bar\phi_{tc})r +i(\phi_c-\bar\phi_c+\phi_{tc}-
\bar\phi_{tc})f$$
$$
+i(\phi_c-\bar\phi_c-\phi_{tc}+
\bar\phi_{tc})\hat f +
{1\over \sqrt{\a '}}(e^{i\rho} d^\a\N_\a(\phi_c+\phi_{tc})\hat\chi+
e^{-i\rho} \bar d^\ad\Nb_\ad(\bar\phi_c+\bar\phi_{tc})\hat{\bar\chi}+
$$
$$e^{i\hat\rho}\hat d^\ah\Nh_\ah(\phi_c+\bar\phi_{tc})\chi+
e^{-i\hat\rho}\hat {\bar d^\ah}\Nbh_\adh(\bar\phi_c+\phi_{tc})\bar\chi)+
S_{aux}]$$
where $\eps_c$ and $\eps_{tc}$
super-worldsheet chiral and twisted-chiral densities\howegris,
$\phi_{\k\hat\k}^c$ and
$\phi_{\k\hat{\bar\k}}^{tc}$
are defined as in the heterotic case, and $S_{aux}$ describes the coupling
of $\phi$ to the worldsheet auxiliary fields.

Although classical
worldsheet supersymmetry
requires that $\P_c=e^{-\phi_c}$ and $\P_{tc}=e^{-\phi_{tc}}$
are chiral and twisted-chiral,
it does not require that they are restricted superfields, i.e. that they
satisfy the reality constraints
$(\N)^2\P_c=(\Nbh)^2\bar \P_c$ and
$(\N)^2\P_{tc}=(\Nh)^2\bar \P_{tc}$. As in the case of the superfields for
the compactification moduli, quantum
$N=(2,2)$ superconformal invariance is expected
to imply these reality conditions (or a suitable modification),
as well as the equations of motion.
Note that $S_{aux}$ vanishes
when $\phi_c +\bar \phi_c=\phi_{tc}
+\bar\phi_{tc}$ (i.e. when $\P_c\bar \P_c=
\P_{tc}\bar \P_{tc}$) since
$d$ couples to $(\phi_c+\bar\phi_c-\phi_{tc}-\bar\phi_{tc})$,
$\psi$ couples to  $\N_\a(\phi_c-\phi_{tc})=\N_\a
(\phi_c+\bar\phi_c-\phi_{tc}-\bar\phi_{tc})$, and $b$ couples to
$\N_\a\Nh_\ah \phi_c=
\N_\a\Nh_\ah
(\phi_c+\bar\phi_c-\phi_{tc}-\bar\phi_{tc})$.
Because $S_{aux}$ violates N=(2,2) superconformal invariance in
a manner which does not appear to be cancelled by anomalies in the
classical term of the sigma model,
$\P_c\bar \P_c=
\P_{tc}\bar \P_{tc}$
is expected to be the on-shell superspace equation
of motion (at least to lowest order in $\a '$).
Note that when $L_{jk}=\d_{jk}$ and all CY
fields are set to zero,
$\P_c\bar \P_c=
\P_{tc}\bar \P_{tc}$ is indeed the equation of motion implied by varying
the weight 2 conformal supergravity
scalar in the low-energy Type II effective action
at the end of section 5.

 From the form of the Type II Fradkin-Tseytlin term, it is reasonable to call
the $\t=\tb=\th=\tbh{}=0$ component of $\phi_c +\bar\phi_c+\phi_{tc}
+\bar\phi_{tc}$ the dilaton, and the
$\t=\tb=\th=\tbh=0$ component of $\phi_c-\bar\phi_c
\pm(\phi_{tc}-\bar\phi_{tc})$ the axions. Just as the dilaton
zero mode couples to the worldsheet Euler number, the axion zero
modes couple to the worldsheet right and left-moving
U(1) instanton number.

As in the heterotic case, one can combine
the Type II Fradkin-Tseytlin term
with the kinetic terms for $\rho$ and
$\hat\rho$ to obtain
\eqn\arb{
\int dz^+ dz^- [\dzp( \rho+i(\phi_c-\bar\phi_c+\phi_{tc}-\bar\phi_{tc}))
D_- (\rho+
i(\phi_c-\bar\phi_c+\phi_{tc}-\bar\phi_{tc}))}
$$
+\dzm(\hat\rho+i(\phi_c-\bar\phi_c-\phi_{tc}+\bar\phi_{tc}))
D_+ (\hat\rho+
i(\phi_c-\bar\phi_c-\phi_{tc}+\bar\phi_{tc}))$$
$$
+(\phi_c+\bar\phi_c+\phi_{tc}+\bar\phi_{tc})r
+{1\over\sqrt{\a '}}
(e^{i\rho} d^\a\N_\a(\phi_c+\phi_{tc})\hat\chi+
e^{-i\rho} \bar d^\ad\Nb_\ad(\bar\phi_c+\bar\phi_{tc})\hat{\bar\chi}+$$
$$
e^{i\hat\rho}\hat d^\ah\Nh_\ah(\phi_c+\bar\phi_{tc})\chi+
e^{-i\hat\rho}\hat {\bar d^\ah}\Nbh_\adh(\bar\phi_c+\phi_{tc})\bar\chi)+
S_{aux}]$$
which is invariant under the spacetime U(1)$\times$U(1) transformations
\eqn\arc{\d\phi_c=\L+\hat\L,~\d\bar\phi_c=-\L-\hat\L,
{}~\d\phi_{tc}=\L-\hat\L,~\d\bar\phi_{tc}=-\L+\hat\L,}
$$\d\Gamma_M=\N_M\L,~\d E_\a{}^M=\L E_\a{}^M,
{}~\d E_\ad{}^M=-\L E_\ad{}^M$$
$$\d\hat\Gamma_M=\N_M\hat\L,~\d E_\ah{}^M=\hat\L E_\ah{}^M,
{}~\d E_\adh{}^M=-\hat\L E_\adh{}^M,$$
$$\d \rho=-4i\L,~\d d^\a=\L d^\a,
{}~\d \bar d^\ad=-\L \bar d^\ad,
{}~\d \hat\rho=2i\hat\L,~\d \hat d^\ah=\hat\L\hat d^\ah,
{}~\d \hat{\bar d}{}^\adh=-\hat\L \hat{\bar d}{}^\adh.$$
Note that $S_{aux}$ is separately invariant under these transformations
since it only depends on the U(1)$\times$U(1) invariant combination
$\phi_c+\bar\phi_c-\phi_{tc}-\bar\phi_{tc}$.

\newsec{ Effective Actions}

The most straightforward method for constructing the effective action is
to use $\b$-function methods to compute the low-energy equations of motion,
and then look for an action which yields these equations. Although
this $\b$-function method is necessary for computing the explicit form
of the action, we can learn a lot just by requiring that the action
has the symmetries predicted by
the sigma model. (By ``effective action'', we always mean the true
effective action, rather than the possibly anomalous Wilson effective
action.)
Since the superfields in the sigma model are conformally gauge-fixed by
$\tilde L=1$ or $L_{jk}=\d_{jk}$, it will be convenient to first
construct superspace actions in this conformal gauge, and then
remove the gauge-fixing condition to obtain conformally-invariant
superspace actions.

\subsec{Heterotic superspace effective actions}

 From the form of the heterotic
sigma model, the dimensionless coupling constant
$\lambda$ (which couples to string loops) can be absorbed in the
effective action by rescaling
the conformal compensator $\P \to \l\P$. This means that string loops
are counted not just by one component field, but by an entire
superfield.

Similarly, the dimensionful coupling constant $\a '$ (which couples to
the classical term in the sigma model) can be absorbed by
rescaling $E_M{}^a\to (\a ')^\half E_M{}^a$
and $E_M{}^\a\to (\a ')^{1\over 4} E_M{}^\a$.
($d^\a$ must also rescale to $(\a ')^{3\over 4} d^\a$ in the
sigma model, but this is just
a redefinition of worldsheet variables.)
Because the effective action is defined to
be invariant under the conformal transformation
$\d E_M{}^a=-\L E_M{}^a$,
$\d E_M{}^\a=-\half\L E_M{}^\a$,
$\d \tilde L=2\L \tilde L$,
$\d \P={3\over 4}\L\P$,
$\a '$ can be absorbed in the conformally-invariant action by rescaling
$\tilde L\to\a ' \tilde L$ and $\P\to (\a ')^{3\over 4}\P$.

In terms of the conformally gauge-fixed superfields which appear in the
heterotic
sigma model, the most general N=1 superspace effective action which
is U(1) invariant and satisfies the
above properties is:
\eqn\arh{\sum_{g=0}^\infty \lambda^{2g-2} \int d^4 x(
\int d^2 \t d^2\tb E^{-1} \sum_{n_I=-\infty}^\infty
 \P^{1-g+n_I} \bar\P^{1-g-n_I} \sum_{p=0}^\infty
(\a ')^{{p-2}\over 2} K_{g,n_I}^p}
$$
+\int d^2\t ~\P^{2-2g} (\a ')^{{3g-3}\over 2} F_g +
\int d^2\tb ~\bar\P^{2-2g} (\a ')^{{3g-3}\over 2} \bar F_g)$$
where $K_{g,n_I}^p=(K_{g,-n_I}^p)^*$ is a general
superfield with U(1) weight $n_I$
and conformal weight $p$, and $F_g$ is a chiral superfield with U(1) weight
$g$ and conformal weight $3g$ (the
conformal weight of an N=1 chiral superfield must be three times its
U(1) weight). We have explicitly separated out the $\P$ and$\bar\P$
zero modes in \arh, so $K_{g,n_I}^p$ and $F_g$ are defined to be
independent of these zero modes.

Note that while the real $d^2\t d^2\tb$ integral is written with
a factor of $E^{-1}=sdet(E_A{}^M)^{-1}$ to make it a density, this real factor
can not appear in the chiral $d^2\t$ integrand. However, in a chiral
basis where $\Nb_\ad=\d_\ad^\md \p_{\ad}$, $\Phi$ acts as the
corresponding density. (We will always be using a chiral basis
when defining chiral integrals.) In fact, all truly chiral superfields are
such densities, with density weight corresponding directly
to conformal weight. This simplifying feature of chiral integrals
is one reason why
their component evaluation is simpler than integrals over full
superspace. A similar procedure can also be applied to non-supersymmetric
theories, with gravity written as conformal gravity plus a compensating
scalar: By writing all fields as densities, all factors of $\sqrt{-g}$
can be removed from the action. This procedure is useful for
string theory, because it is the density form of the dilaton field,
with the same weight as $(-g)^{{1\over 4}}$, that is invariant under
T-duality. For example, the low-energy action of the bosonic closed
string can be written with this dilaton field as the only density.

The effective action of \arh can be easily written in conformally-invariant
form by simply inserting an appropriate power of $\tilde L=$
$ L+\a '(c_1 \Omega_{YM}+c_2 \Omega_{Lorentz})$ and removing the conformal
gauge-fixing condition $\tilde L=1$. Since the non-chiral measure
carries conformal weight $-3$, the chiral measure carries
conformal weight $-2$, ${\Phi}$ carries conformal weight ${{3}\over
2}$, and $\tilde L$ carries conformal weight $+2$,
the conformally-invariant form of \arh is
\eqn\arj{\sum_{g=0}^\infty \lambda^{2g-2} \int d^4 x}
$$(
\int d^2 \t d^2\tb  E^{-1}\sum_{n_I=-\infty}^\infty
 \P^{1-g+n_I} \bar\P^{1-g-n_I} \sum_{p=0}^\infty
(\a ')^{{p-2}\over 2} \tilde L^{{3g-1-p}\over 2} K_{g,n_I}^p$$
$$
+\int d^2\t ~\P^{2-2g} (\a ')^{{3g-3}\over 2} F_g +
\int d^2\tb ~\bar\P^{2-2g} (\a ')^{{3g-3}\over 2} \bar F_g).$$
(Note that $K_{g,n_I}^p$ may involve scale-invariant functions
of $\tilde L$ such as $\p_m (\log\tilde L)$.)

Since the U(1) weight $g$ of $F_g$ must be absorbed by $\P^{2-2g}$,
$F_g$ can only occur with coupling $\lambda^{2g-2}$.
This means that $F_g$ cannot receive perturbative
or non-perturbative quantum
corrections. In other words,
after expanding $\Phi$ and $\tilde L$ around their
vacuum expectation values (which can be set to one by rescaling
$\l$ and $\a '$), $F_g$
appears only once in the perturbative
expansion. Of course, this does not prevent quantum corrections if
a chiral $F$-term can also be written as a non-chiral $D$-term, which
could then receive quantum corrections through
U(1)-neutral factors of $\Phi\bar\Phi$.

Note that any chiral $F$-term can be written as a non-chiral $D$-term
if one allows the non-local operator $(\N)^2/\bo$.
However, non-local $1/\bo$ terms in the effective action come
from anomalies, which are not expected to receive quantum
corrections\ref\anom{I. Antoniadis, K. Narain and T. Taylor, Phys. Lett.
267B (1991) 37\semi M. Bershadsky, S. Cecotti, H. Ooguri and C. Vafa,
Comm. Math. Phys. 165 (1994) 311.}.
Furthermore, some chiral $F$-terms can be written as
local non-chiral $D$-terms
by pulling a factor of $(\Nb)^2$ off a field-strength. Normally, this
would not lead to quantum corrections, since adding $\P\bar\P$ to
the $D$-term would break gauge invariance. However, in the heterotic
effective action where the tensor multiplet transforms anomalously under
Lorentz and Yang-Mills gauge transformations, this type of $D$-term
can sometimes lead to quantum corrections.

For example, the $F$-term
\eqn\ark{\lambda^{2g-2} {\a '}^{{3g-3}\over 2} \int d^4 x \int d^2 \t
{}~\P^{2-2g}
(W^I_\a W^{\a I})^g \tau_g(M^{(i)})}
can also be written as the  non-chiral $D$-term
\eqn\arl{\lambda^{2g-2} {\a '}^{{3g-3}\over 2} \int d^4 x\int
d^2 \t d^2\tb E^{-1}~\Phi^{2-2g}(W^I_\a W{^\a I})^{g-1}
\Omega_{YM} ~\tau_g(M^{(i)})}
where $(\Nb)^2\Omega_{YM}=W_\a^I W^{\a I}$
and $\tau_g(M^{(i)})$ is a chiral function of the compactification moduli.
This non-chiral $D$-term can get corrections, for example, from
\eqn\arm{\lambda^{2(g+n)-2} {\a '}^{{3g-3}\over 2}
\int d^4 x\int d^2\t d^2\tb }
$$E^{-1}~\Phi^{2-2g-n}\bar\Phi^{-n} (W_\a W^\a)^{g-1}
\Omega_{YM} \tilde \tau_g(M^{(i)},\bar M^{(i)})
{\tilde L}^{{3n}\over 2},$$
which would give quantum corrections to $\tau_g$ of $\l^{2n}
 \tilde\tau_g(
M^{(i)},\bar M^{(i)}).$
So the $F$-term is not protected against quantum corrections. Note, however,
that because \arm is not gauge-invariant, it must occur in the
combination
\eqn\arn{\lambda^{2g-2} {\a '}^{{3g-5}\over 2}\int d^4 x
\int d^2\t d^2\tb E^{-1} \P^{2-2g-n}\bar\P^{-n}}
$$(W_\a W^\a)^{g-1}
(L+\a '(c_1\Omega_{YM} +c_2\Omega_{Lorentz}))
\tilde\tau_g(M^{(i)},\bar M^{(i)})
\tilde L^{{3n} \over 2}.$$
This means that quantum corrections to the $F$-term of \ark
 are related by
the proportionality constant $c_1/c_2$ to quantum corrections of the $F$-term
\eqn\arp{\lambda^{2g-2} {\a '}^{{3g-3}\over 2}\int d^4 x \int d^2\t
{}~\P^{2-2g}
W_{\a\b\g} W^{\a\b\g} (W_\d W^\d)^{g-1} \tau_g(M^{(i)}).}

At low energies, only $K_{0,0}^0$ contributes to the non-chiral
$D$-term in the tree-level effective action.
($n_I=0$ since U(1)-charged superfields
have non-zero conformal weight.)
Since only the compactification moduli
have conformal weight zero, $K_{0,0}^0$
must be some function $K_0 (M^{(i)},\bar M^{(i)})$.
The tree-level low-energy effective action is therefore
\eqn\arq{\lambda^{-2}{\a '}^{-1}
 \int d^4 x[\int d^2\t
d^2 \tb ~ \Phi\bar\Phi(L+\a ' (c_1 \Omega_{YM}+c_2\Omega_{Lorentz}))^{-\half}
 K_0(M^{(i)},\bar M^{(i)}) +}
$${\a '}^{-\half}\int d^2\t ~\P^2 F_0(M^{(i)})+
{\a '}^{-\half}\int d^2\tb ~\bar \P^2 \bar F_0(\bar M^{(i)})].$$
The supergravity, tensor, and super-Yang-Mills actions come from
the $D$-term, while the cosmological constant and Yukawa couplings
come from the $F$-term.
Note that the kinetic term for Yang-Mills fields gets tree-level
contributions from the $D$-term $\l^{-2}\int d^4 x $
$\int d^2\t d^2\tb$
$
c_1 \Omega_{YM} K(M^{(i)},\bar M^{(i)})$ and gets one-loop
contributions from the $F$-term
$\int d^4 x $
$\int d^2\t $
$W^I_\a W^{\a I} F_1(M^{(i)})$.
Although many aspects of the heterotic superspace effective action have
already been discussed in reference \cec,
the Type II superspace effective
action has not yet appeared in the literature.

\subsec { Type II superspace effective action }

 From the form of the Type II
sigma model, the dimensionless coupling constant
$\lambda$ (which couples to string loops) can be absorbed in the
effective action by rescaling
the chiral and twisted-chiral
compensators $\P_c \to \l\P_c$ and $\P_{tc}\to\l\P_{tc}$. As explained in
section 3, $\P_c$ can be identified with a vector field strength $W^{(0)}$
and $\P_{tc}$ can be identified with a tensor field strength $L_{--}^{(0)}$.
Therefore, string loops in the Type II effective action are counted
by the superfields $W^{(0)}$ and $L_{jk}^{(0)}$. (This loop-counting
assumes that all compactification moduli have been expressed in terms of
the dimensionless superfields $M_c^{(i)}$ and $M_{tc}^{(i)}$.)
Since the graviphoton field strength is one of the components of $W^{(0)}$,
its kinetic term (which is quadratic in $W^{(0)}$) appears at order
$\l^{-2}$, as expected for a tree-level contribution.

As in the heterotic sigma model,
the dimensionful coupling constant $\a '$
can be absorbed by
rescaling $E_M{}^a\to (\a ')^\half E_M{}^a$
and $E_M{}^{\a j}\to (\a ')^{1\over 4} E_M{}^{\a j}$.
Because the effective action is defined to
be invariant under the conformal transformation
$\d E_M{}^a=-\L E_M{}^a$,
$\d E_M{}^{\a j}=-\half\L E_M{}^{\a j}$,
$\d L_{jk}=2\L L_{jk}$,
$\d W^{(0)}=\L W^{(0)}$,
$\d L_{jk}^{(0)}=2\L L_{jk}^{(0)}$,
$\a '$ can be absorbed in the conformally-invariant action by rescaling
$L_{jk}\to \a ' L_{jk}$, $W^{(0)} \to (\a ')^{\half}
W^{(0)}$, and
$L^{(0)}_{jk} \to \a ' L^{(0)}_{jk}$.

For the Type II superstring, mirror symmetry further restricts the
superspace effective action. Mirror transformations relate Type IIA
compactifications to Type IIB compactifications by switching $\hat G_C$
with $\hat{\bar G}_C$ (the right-moving fermionic generators for the
compactification
N=2) and switching
$\Omega^{(i)}_c$ with $\Omega^{(i)}_{tc}$ (the worldsheet
chiral and twisted-chiral primary fields).
 From the form of the Type II sigma model of \alb and \arb,
this switch can be
undone by switching $E_M{}^{\a -}$ with $E_M{}^{\adh +}$, $\P_c$ with
$\P_{tc}$, and $M_c^{(i)}$ with $M_{tc}^{(i)}$ (one must also switch
$\hat\rho$ with $-\hat\rho$,
$\hat d_\ah$ with $\hat{\bar d}_\adh$, and $\Pi^0_+$ with
$-\Pi^0_+$, but this is just a redefinition of
worldsheet variables). So after gauge-fixing $L_{jk}=\d_{jk}$ and
imposing the standard reality conditions on $\P_c$ and $\P_{tc}$,
mirror symmetry relates Type IIA effective actions to Type IIB
effective actions by switching $E_M{}^{\ah -}$ with $E_M{}^{\adh +}$,
$W^{(0)}$ with
$L_{--}^{(0)}$, $\bar W^{(0)}$ with
$L_{++}^{(0)}$, and $M_c^{(i)}$ with $M_{tc}^{(i)}$.
(Note that
$E_M{}^{\ah -}$ and $E_M{}^{\adh +}$ will always appear in
combinations which allow this shift to preserve Lorentz invariance.)

Furthermore, mirror symmetry implies that the tensor multiplets
$L^{(0)}_{jk}$
and $L^{(i)}_{jk}$ only appear through $\P_{tc}$ and $M_{tc}^{(i)}$, and
therefore, $L^{(0)}_{+-}$ and $L^{(i)}_{+-}$ never appear explicitly in the
sigma model. Although
$L^{(0)}_{+-}$ and $L^{(i)}_{+-}$ appear implicitly in
$L^{(0)}_{++}$ and $M^{(i)}_{tc}$ through the linear constraint, their lowest
component, $l^{(0)}_{+-}$ and $l^{(i)}_{+-}$,
always appears with derivatives.
Therefore, after gauge-fixing $L_{jk}=\d_{jk}$,
the Type II superspace effective action
must be invariant under the Pecci-Quinn-like shifts
\eqn\are{\d L^{(0)}_{+-}=c^{(0)},\quad
\d L^{(i)}_{+-}=c^{(i)} ,}
where $c^{(0)}$
and $c^{(i)}$ are independent constants. This implies that before
gauge-fixing $L_{jk}$, the Type II effective action must be invariant under
\eqn\arf{\d L^{(0)}_{jk}=c^{(0)} L_{jk},\quad
\d L^{(i)}_{jk}=c^{(i)} L_{jk},}
which is the unique conformally and SU(2)-invariant generalization of \are.

In terms of the conformally and SU(2)/U(1)
gauge-fixed superfields appearing in the Type II sigma model,
the most general N=2 superspace action which is U(1)$\times$U(1)
invariant and satisfies the above properties
is:
\eqn\asc{\sum_{g=0}^\infty \lambda^{2g-2} \int d^4 x}
$$
(\int d^2 \t d^2\tb d^2\th d^2\tbh E^{-1}
\sum_{(n_I,\hat n_I, s)=-\infty}^\infty
 \P_c^{\half(1-g+n_I+\hat n_I+s)} \bar\P_c^{\half(1-g-n_I-\hat n_I +s)}$$
$$
 \P_{tc}^{\half(1-g+n_I-\hat n_I-s)} \bar\P_{tc}^{\half(1-g-n_I+\hat n_I -s)}
\sum_{p=0}^\infty
(\a ')^{p\over 2} K_{g,n_I,\hat n_I,s}^p$$
$$
+\int d^2\t d^2\th~\P_c^{2-2g} (\a ')^{g-1} F^c_g +
\int d^2\tb d^2\tbh ~\bar\P_c^{2-2g} (\a ')^{g-1} \bar F^c_g$$
$$
+\int d^2\t d^2\tbh~\P_{tc}^{2-2g} (\a ')^{g-1} F^{tc}_g +
\int d^2\tb d^2\th~\bar\P_{tc}^{2-2g} (\a ')^{g-1} \bar F^{tc}_g)$$
where $K_{g,n_I,\hat n_I,s}^p=(K_{g,-n_I,-\hat n_I,s}^p)^*$ is a general
superfield with U(1)$\times$U(1) weight $[n_I+\hat n_I, n_I-\hat n_I]$
and conformal weight $p$ ($s$ is unrestricted),
$F^c_g$ is a chiral superfield with U(1)$\times$U(1) weight
$[2g,0]$ and conformal weight $2g$ (the
conformal weight of an N=2 chiral superfield must be equal its
U(1) weight),
and $F^{tc}_g$ is a twisted-chiral superfield with U(1)$\times$U(1) weight
$[0,2g]$ and conformal weight $2g$ (although the
conformal weight of an N=2 twisted-chiral superfield is unrestricted
when $L_{jk}$ is gauge-fixed, mirror symmetry forces $F^{tc}_g$ to have
the same conformal weight as $F^c_g$).
We have explicitly separated out the $\P$
zero modes in \asc, so $K_{g,n_I,\hat n_I,s}^p$,
$F^c_g$, and $F^{tc}_g$ are defined to be
independent of these zero modes (however, they can depend on the
moduli $M_c^{(i)}=\log(W^{(i)}/\P_c)$ and
$M_{tc}^{(i)}=\log(L_{--}^{(i)}/\P_{tc})$).

As in the heterotic superstring, $F_g^c$ and $F_g^{tc}$ do not receive
perturbative quantum
corrections since
they must appear in the effective action only at order $\l^{2g-2}$
so that their U(1)$\times$U(1) charge is compensated by the
U(1)$\times$U(1) charge of $\P_c$ or $\P_{tc}$. In other words, after
expanding $\P_c$ and $\P_{tc}$ around their vacuum expectation values,
$F_g^c$ and $F_g^{tc}$ appear only once in the perturbative expansion.
Furthermore, mirror symmetry implies that
$F_g^c$ for Type IIA compactifications is related to
$F_g^{tc}$ for Type IIB compactifications
by switching $E_M{}^{\ah -}$ with $E_M{}^{\adh +}$ and $M_c^{(i)}$ with
$M_{tc}^{(i)}$.

We are assuming that the chiral and twisted-chiral $F$-terms in
\asc can not be written as non-chiral $D$-terms, which could
then receive quantum corrections from U(1)-neutral combinations
of $\P_c$ and $\P_{tc}$. (See the previous
subsection for why such $D$-terms are unlikely.) Note that if a twisted-chiral
$F$-term can be written as such a $D$-term, then mirror symmetry implies
that the mirror chiral $F$-term can also be written as such a $D$-term.
So in the unlikely event that
$F_g^{tc}$ receives perturbative corrections, so does $F_g^c$.
However, as will be discussed at the end of this section, if mirror
symmetry were non-perturbatively broken, $F_g^{tc}$
might receive non-perturbative corrections without $F_g^c$
receiving corrections.

One type of $g$-loop chiral and twisted-chiral term
is
\eqn\ase{\l^{2g-2} (\a ')^{g-1}\int d^4 x
[\int d^2\t d^2\th~\P_c^{2-2g} (P_{\a\bh}P^{\a\bh})^{g} \tau^c_g (M_c^{(i)})
+ c.c.}
$$+\int d^2\t d^2\tbh~\P_{tc}^{2-2g} (Q_{\a\bdh}Q^{\a\bdh})^g
\tau^{tc}_g
(M_{tc}^{(i)})+ c.c.]$$
where $P_{\a\b}$ and $Q_{\a\bd}$ are the chiral and twisted-chiral
supergravity field strengths which appear in the sigma
model of \alb. These terms describe the scattering of two gravitons with
either $2g-2$ graviphotons or $2g-2$ hypermultiplets, and
occur only at $g$ string-loops
in the perturbative $S$-matrix
\ref\top
{I. Antoniadis, E. Gava, K. Narain and T. Taylor, Nucl. Phys. B413
(1994) 162.}\metop.
$\tau^c_g$ and $\tau^{tc}_g$ are functions which depend only on topological
properties of the compactification manifold, and since mirror symmetry
exchanges
$P_{\a\bh}P^{\a\bh}$ with
$Q_{\a\bdh}Q^{\a\bdh}$,
$\tau^c_g$ for Type IIA compactifications is equal to $\tau^{tc}_g$
for Type IIB compactifications, and
$\tau^{tc}_g$ for Type IIA compactifications is equal to $\tau^{c}_g$
for Type IIB compactifications.

By removing the gauge-fixing condition on $L_{jk}$, the non-chiral and
chiral terms
in \asc can easily
be written in conformally and SU(2)-invariant form as
\eqn\asg{\sum_{g=0}^\infty \lambda^{2g-2} \int d^4 x(
\int d^2 \t d^2\tb d^2\th d^2\tbh E^{-1}
\sum_{(n_I,\hat n_I, s)=-\infty}^\infty}
$$(W^{(0)})^{\half(1-g+n_I+\hat n_I+s)}
(\bar W^{(0)})^{\half(1-g-n_I-\hat n_I +s)}$$
$$
(y^j y^k L^{(0)}_{jk})^{\half(1-g+n_I-\hat n_I-s)}
(\bar y^l \bar y^m L^{(0)}_{lm})^{\half(1-g-n_I+\hat n_I -s)} $$
$$
\sum_{p=0}^\infty
(\a ')^{p\over 2}(L^{jk}L_{jk})^{{1\over 4}(3g-3+s-2p)}
K_{g,n_I,\hat n_I,s}^p (y,\bar y)$$
$$
+\int d^4\t (W^{(0)})^{2-2g} (\a ')^{g-1} F^c_g +
\int d^4\tb  (\bar W^{(0)})^{2-2g} (\a ')^{g-1} \bar F^c_g)$$
where $y^j \bar y^k$ is defined as $\epsilon^{jk} +{L^{jk}\over
\sqrt{L_{lm} L^{lm}}}$
in the non-chiral integrand, and $K_{g,n_I,\hat n_I,s}^p (y,\bar y)$
is defined by contracting all SU(2) $+$ indices with $y$'s and
all SU(2) $-$ indices
with $\bar y$'s. Note that there are an equal number of $y$'s and
$\bar y$'s in the integrand because of U(1) invariance.

However, in order to write the twisted-chiral term of \asc in conformally
and SU(2)-invariant form, one needs to introduce two independent
complex variables,
$u^+$ and $u^-$\harm. Using $u^\pm$ and any linear superfield
$h_{j_1 ... j_n}$ satisfying
\eqn\asl{\N_{\a (j_0}h_{j_1 ... j_n)}=
\Nb_{\ad (j_0}h_{j_1 ... j_n)}=0,}
one can define a ``harmonic'' superfield  of
degree $n$, $\tilde h(u^+,u^-)=h_{j_1 ... j_n} u^{j_1} ... u^{j_n}$,
which satisfies
\eqn\asm{u^j \N_{\a j} \tilde h=
u^j \Nb_{\ad j}\tilde h=0.}
Note that the product of two harmonic superfields of degree $n_1$ and
$n_2$ is a harmonic superfield of degree $n_1+n_2$.

Using a harmonic superfield $\tilde h$
of degree 2 (which has conformal weight
2 by N=2 superspace rules), one
can then define the conformally and SU(2)-invariant action\roc
\eqn\asq{\int d^4 x \oint u_j du^j \int d^4\t_\natural \tilde h
=\int d^4 x\oint u_j du^j {{\int (v^j d \t_j)^2 \int (v^k d \tb_k)^2 }
\over {(v^l u_l)^4}} ~
\tilde h}
where $v^j$ is arbitrary (because $\tilde h$ is harmonic, the integral
is independent of $v^j$) and
$\oint du$ is defined as a contour integral in $CP1$
around a pole of $\tilde h$. (Note that the action is invariant under the
complex projective transformation $u^j \to q u^j$.)

In order to reproduce the twisted-chiral term of \asc, $\tilde h$ should
be defined as
\eqn\atc{\tilde h=\lambda^{2g-2}(\a ')^{g-1}
{(\tilde L^{(0)})^{2-2g} \tilde F_g\over \tilde L}}
where
$\tilde L^{(0)}= u^j u^k L^{(0)}_{jk}$,
$\tilde L= u^j u^k L_{jk}$, and
$\tilde F_{tc}$ is a harmonic superfield of degree $4g$ which
satisfies $\tilde F_g(u^-=1,u^+=0)=F^{tc}_g$
when $L_{jk}$ is gauge-fixed to $\d_{jk}$.
(Note that $\N_- \tilde F_g=\Nb_-\tilde F_g=0$ when $u^-=1$ and $u^+=0$, so
$F^{tc}_g$ is a twisted-chiral superfield of U(1)$\times$U(1) charge
$[0,2g]$.) With this choice of $\tilde h$, \asq is
\eqn\afy{
\lambda^{2g-2}(\a ')^{g-1}
\int d^4 x \oint u_j du^j \int d^4\t_\natural
{(\tilde L^{(0)})^{2-2g} \tilde F_g\over \tilde L}}
where the contour
integral is performed around the pole of
$\tilde L=u^j u^k L_{jk}$.

When $L_{jk}$ is gauge-fixed to $\d_{jk}$,
the action of \afy becomes
\eqn\atf{
\lambda^{2g-2}(\a ')^{g-1}
\int d^4 x \oint u_k du^k \int d^4 \t_\natural
{(\tilde L^{(0)})^{2-2g} \tilde F_g\over {u^+u^-}}}
$$=
\lambda^{2g-2}(\a ')^{g-1}
\int d^4 x\int d^2 \t d^2 \tbh$$
$$\oint d\xi
{(\tilde L^{(0)})^{2-2g}
(u^-=1,u^+=\xi) \tilde F_g(u^-=1,u^+=\xi)\over {\xi}}$$
$$=
\lambda^{2g-2}(\a ')^{g-1}
\int d^4 x\int d^2 \t d^2 \tbh
(L^{(0)}_{--})^{2-2g} F^{tc}_g$$
where we have chosen the $CP1$ basis $(u^-,u^+)=(1,\xi)$ and performed
a contour integral around $\xi=0$.
Since $\Phi_{tc}=L_{--}^{(0)}$, \afy and its complex
conjugate reproduce the
twisted-chiral and twisted-anti-chiral terms of \asc.

Note that \afy is invariant under the Pecci-Quinn-like shifts
$$\d L^{(0)}_{jk}=c^{(0)} L_{jk},\quad \d L^{(i)}_{jk}=c^{(i)} L_{jk}$$
where $c^{(0)}$ and $c^{(i)}$ are independent constants, since
these shifts eliminate the pole when $\tilde L$ vanishes.
However, it is plausible that this shift symmetry is broken by
spacetime non-perturbative instantons, just as other types of
Pecci-Quinn symmetries can be broken. (This would
imply that mirror symmetry is broken non-perturbatively since
mirror symmetry implies that linear superfields appear only through
$\P_{tc}$ and $M_{tc}^{(i)}$ in the effective action.)

If the shift symmetry were non-perturbatively broken, one could
consider harmonic actions of the type
\eqn\awu{
\lambda^{2g-2}(\a ')^{g-1}
\int d^4 x\oint u_j du^j \int d^4 \t_\natural
(\tilde L^{(0)})^{2-2g} y({\tilde L^{(0)}\over {\lambda\tilde L}})
{\tilde f \over {\tilde L}}}
where $\tilde f=f_{j_1 ... j_{4g}}u^{j_1} ... u^{j_{4g}}$
is a harmonic superfield of degree $4g$ and $y(
{\tilde L^{(0)} /{\lambda\tilde L}})$ is an arbitrary function (e.g.,
$y({\tilde L^{(0)}/ {\lambda\tilde L}})=\exp(-
{\tilde L^{(0)}/ {\lambda\tilde L}})$).

Expanding around the background expectation value of
$\tilde L^{(0)}$/$\tilde L$, this term would
give corrections to hypermultiplet interactions of the form
$\d F^{tc}_g = y({1\over\lambda}) f_{- ... -}$.
Note, however, that vector multiplet
interactions can not receive non-perturbative corrections
since, as in the heterotic superstring effective action, the linear
superfield can not appear in a chiral action.

Finally, the tree-level low-energy effective action for the
Type II superstring is given by
\eqn\avv{
{1\over{\lambda^2 \a '}}
\int d^4 x
[\int d^4\t  (W^{(0)})^2 F^c_0 (M_c^{(i)}) +
c.c.}
$$+
\oint u_j du^j \int d^4 \t_\natural
{(\tilde L^{(0)})^2\over \tilde L} F^{tc}_0 (\tilde M_{tc}^{(i)}) + c.c.]$$
where $\tilde M_c^{(i)}=
W^{(i)}/W^{(0)}$ and $\tilde M_{tc}^{(i)} =\tilde L^{(i)}/\tilde L^{(0)}$.
Although the chiral part of this low-energy action agrees with
the component action of \dewit, the ``harmonic" part of \avv
allows more general tensor hypermultiplet couplings than the
``improved tensor" action of \dewit. Note that unlike the improved
tensor action of \dewit, \avv contains the Pecci-Quinn-like symmetry of
\arf.

Although N=2 supersymmetry also allows the cosmological term
\eqn\axx{{1\over{\lambda^2 \sqrt{\a '}}}
\int d^4 x
\int d^2\t d^2\th d^2\tb d^2 \tbh E^{-1} L_{jk}^{(0)} V^{jk(0)}}
where $V^{jk(0)}$ is the prepotential of the vector compensator,
this term breaks the Pecci-Quinn-like symmetry and is
therefore perturbatively forbidden.

\newsec { Conclusions}

In this paper, we have constructed manifestly spacetime supersymmetric
sigma models and effective actions for $4D$ compactifications of
heterotic and Type II superstrings. For the heterotic superstring,
the sigma model can be found in equations \ala and \awa, and the effective
action in \arj and \arq. For the Type II superstring, the sigma model can
be found in \alb and \arb, and the effective action in \asg, \afy, and
\avv.

We have also proven various non-renormalization theorems for the
superspace effective action, including the theorem that chiral
$F$-terms receive no perturbative or non-perturbative corrections.
For the Type II superstring, mirror symmetry implies that twisted-chiral
$F$-terms (which describe hypermultiplet interactions) are also
unrenormalized. However, it is plausible that mirror symmetry
is non-perturbatively broken by spacetime instantons, which would
allow hypermultiplet interactions to receive non-perturbative corrections.

Our results in this paper were
based on the observation that string loops
in the $4D$ heterotic superstring are counted by an N=1 chiral
compensator, and
string loops in the $4D$ Type II superstring are counted by an N=2
vector compensator and an N=2 tensor compensator. This explains the
coupling of Ramond-Ramond fields, and contradicts the
standard folklore that Type II string loops are counted by just
a hypermultiplet\second. The mistaken folklore was caused
by confusing the physical scalar (which sits in a hypermultiplet and
couples like the determinant of the metric) with the dilaton
compensator (which couples to the worldsheet curvature).

Based on a
similar incorrect reasoning, the standard
folklore also claims that for the $4D$ heterotic superstring
with $N=2$ spacetime supersymmetry,
string loops are counted by a vector multiplet\second.
Although we are presently unsure of the precise supersymmetry multiplet
for the dilaton compensator in this N=2 heterotic case, we believe that
it is not just a vector multiplet, and therefore the
standard folklore is again incorrect.

An obvious question is if our techniques can be generalized to
the ten-dimensional uncompactified superstring. Although there does
exist an N=2 worldsheet-supersymmetric description of the uncompactified
superstring, this description is not manifestly SO(9,1) invariant\meold.
This makes it difficult to extend the worldsheet action in a
flat superspace background to a sigma model action in a curved
superspace background.
Nevertheless, we conjecture that if such an extension were performed
(probably using harmonic variables to make Lorentz invariance manifest),
the $10D$ supergravity theory would contain a spacetime conformal
compensator which couples to worldsheet supercurvature in the sigma model.
Hopefully, a proper understanding of this compensator will lead to
an off-shell superspace description of $10D$ supergravity.

A less ambitious question is if our techniques can be generalized to
six-dimensional compactifications of the superstring (which, after toroidal
compactification, would
be useful for understanding the
N=2 $4D$ heterotic superstring). Since
there already exists an N=2 worldsheet-supersymmetric description of the
superstring with manifest SO(5,1) invariance\metop, the answer
is probably yes.
As was shown with Cumrun Vafa, this six-dimensional superstring actually
contains N=4 worldsheet supersymmetry, which can be made manifest by
introducing SU(2)/U(1) harmonic variables. It is likely that, as in
references \ref\harsig{F. Delduc and
E. Sokatchev, Class. Quant. Grav. 9 (1992) 361.}
and \harm, these harmonic variables will be useful for
constructing manifestly spacetime supersymmetric sigma models and
effective actions. A $6D$ superspace effective action
would be very useful for studying the
recent string-duality conjectures which relate the $6D$ heterotic
and Type II superstrings\HT\Wittdim\second.
\vskip 20pt

\centerline{\bf Acknowledgements}

We would like to thank J. de Boer, J. Derendinger, B. de Wit, M. Grisaru,
J. Louis, P. van Nieuwenhuizen, H. Ooguri, M. Ro\v cek,
Ergin Sezgin, Kostas Skenderis, Kellogg Stelle, and Cumrun Vafa for
useful conversations. N.B. would also like to thank the ITP at Stony Brook
for its hospitality. This work was financed by the Conselho Nacional de
Pesquisa and by the National Science Foundation grant No.\ PHY9309888.

\listrefs
\end